\documentclass[aps,twocolumn,superscriptaddress,showpacs,floatfix,longbibliography,nobibnotes]{revtex4-1}
\usepackage{amsmath,amssymb,amsfonts}
\usepackage{color}
\usepackage[pdftex]{hyperref,graphicx}
\hypersetup{colorlinks = true, urlcolor = blue, linkcolor = blue, citecolor = blue}
\usepackage{wscmath}

\newcommand{\expect}[1]{\langle {#1} \rangle}

\begin{document}
\title{Spin-orbit coupled bosons in one dimension:\\  emergent gauge field and Lifshitz transition}
\author{William S. Cole}
\author{Junhyun Lee }
\author{Khan W. Mahmud}
\author{Yahya Alavirad}
\affiliation{
Department of Physics, Condensed Matter Theory Center and Joint Quantum Institute, University of Maryland, College Park, MD 20742
}
\author{I.~B.~Spielman}
\affiliation{Joint Quantum Institute, National Institute of
Standards and Technology, and University of Maryland, Gaithersburg, MD 20899}
\author{Jay D. Sau}
\affiliation{
Department of Physics, Condensed Matter Theory Center and Joint Quantum Institute, University of Maryland, College Park, MD 20742
}

\date{\today}
\begin{abstract}
In the presence of strong spin-independent interactions and spin-orbit coupling, we show that the spinor Bose liquid confined to one spatial dimension undergoes an interaction- or density-tuned quantum phase transition similar to one theoretically proposed for itinerant magnetic solid-state systems. The order parameter describes broken $Z_2$ inversion symmetry, with the ordered phase accompanied by non-vanishing momentum which is generated by fluctuations of an emergent dynamical gauge field at the phase transition. This quantum phase transition has dynamical critical exponent $z \simeq 2$, typical of a Lifshitz transition, but is described by a nontrivial interacting fixed point. From direct numerical simulation of the microscopic model, we extract previously unknown critical exponents for this fixed point. Our model describes a realistic situation of 1D ultracold atoms with Raman-induced spin-orbit coupling, establishing this system as a platform for studying exotic critical behavior of the Hertz-Millis type.
\end{abstract}
\maketitle

%%%%%%%%%%%%%%%%%%%%%%%%%%%%%%%%%%%%%%%%%%%%%%%%%%

\seclabel{intro}

Perhaps the first example of a quantum phase transition (QPT) was Stoner's identification of a zero-temperature critical point distinguishing between unpolarized and spin-imbalanced Fermi liquids, and magnetic transitions in Fermi liquids have remained a rich subject since. These transitions for gapless, itinerant magnets belong to a class that is qualitatively distinct from transitions between gapped phases of matter and still remain mysterious despite the seminal works by Hertz and Millis~\cite{hertz, millis, sachdev_book}. While quantum criticality in Fermi liquids is a pervasive phenomenon in strongly correlated phases of matter, simple realizations of the ferromagnetic transition appear to be rare, especially in low-dimensional fermion systems where there is hope of a detailed theoretical understanding~\cite{kun_yang_fm_PRL, kozii_2017}. Nonetheless, a different paradigm for itinerant ferromagnetism appears if our initial degrees of freedom are bosons. In fact, since the ground state of an interacting (weak or strong) spinor Bose gas is a spin-polarized superfluid (SF)~\cite{eisenberg_lieb_PRL}, ultracold bosons already provide a more natural realization of an itinerant ferromagnetic liquid compared to electrons in solid state systems requiring a Stoner instability. This work explores spin dynamics arising from the interplay of spin-charge separation, a concomitant emergent gauge field, and spin-momentum locking near the ferromagnetic QPT of a spin-orbit coupled interacting 1D Bose liquid \cite{higbie_PRL_2002, lin_Nature_2011_expt, ho_PRL_2011, li_PRL_2012, zhang_PRL_2012_expt, beeler_Nature_2013_expt}.

\begin{figure}[h!]
\centering
	\includegraphics[width=0.46\textwidth]{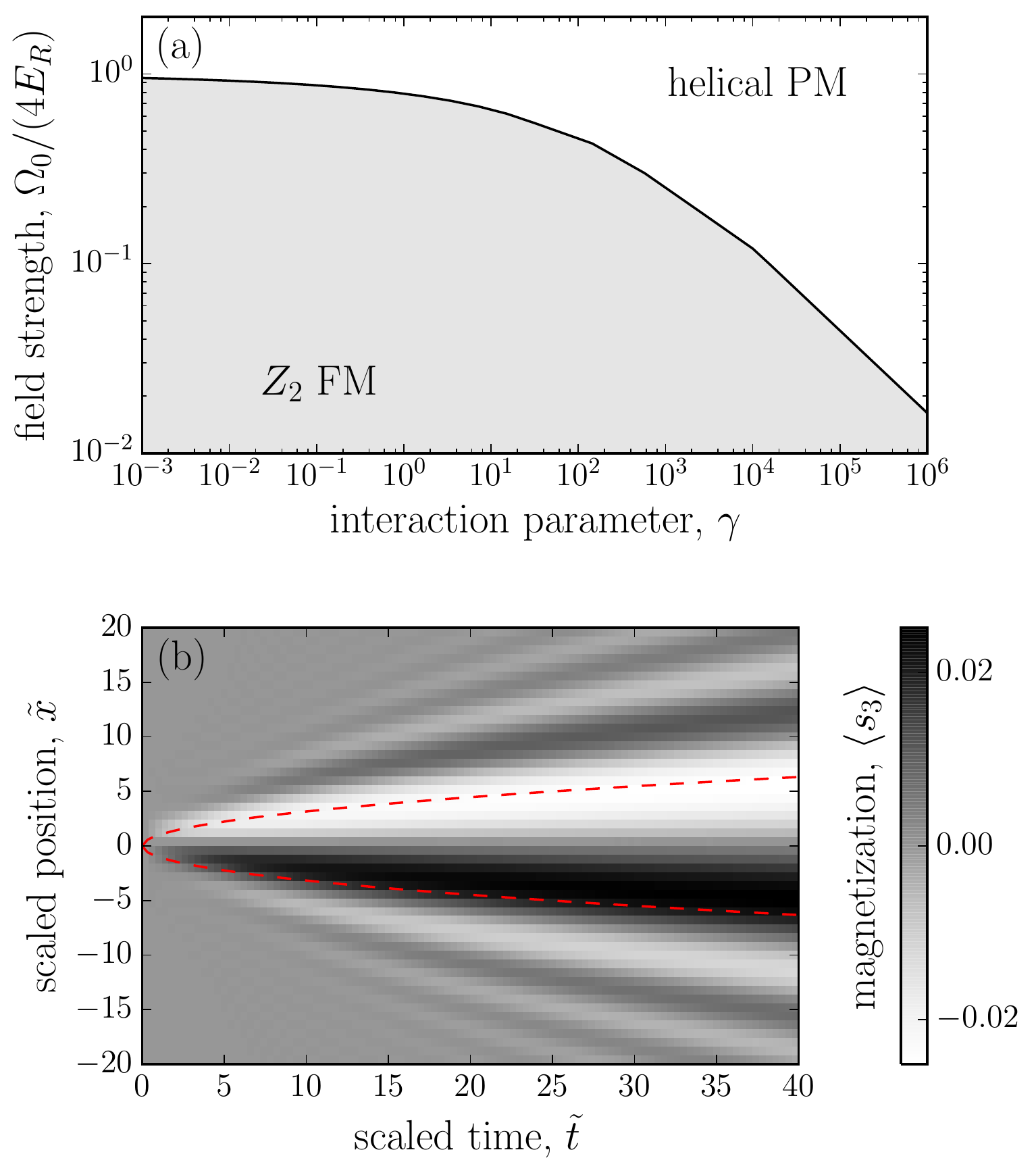}
	\caption{(a) Phase diagram as derived from our effective Lagrangian approach. The critical Zeeman coupling $\Omega_0$ separating the $Z_2$ ordered phase from the disordered phase depends strongly on the dimensionless interaction strength. (b) Evolution of a localized spin fluctuation.  The time evolution was computed using the semiclassical equations of motion \eref{eom} on the critical curve. The initially localized fluctuation spreads as $x \propto t^{1/2}$, consistent with an expected dynamical critical exponent $z \simeq 2$.}
	\flabel{phases}
\end{figure}

The strongly interacting Bose liquid \emph{without} spin-orbit coupling can be well understood by separating the excitations into a quadratically dispersing (but gapless) spin degree of freedom and a massless acoustic mode. Spin-charge separation in itinerant \emph{fermion} ferromagnets has been profitably formulated in terms of emergent gauge fields, for example in the context of solid-state spintronics~\cite{barnes_PRL_2005}. Likewise, in the Bose liquid, fluctuations of the spin degree of freedom behave as an emergent dynamical gauge field for the SF sound mode, the former coupled to the latter by an emergent electric field, as we show in this work. However, without any spin-dependent perturbations, the ferromagnetic ground state is also a fully-polarized spin eigenstate; therefore, in the absence of spin fluctuations the emergent field vanishes.

Spin fluctuations can be induced in the otherwise static spin-polarized gas by the addition of a helical Zeeman field. Qualitatively, a sufficiently strong field polarizes the local magnetization to be entirely parallel to it. However, the spin stiffness of the ferromagnetic liquid leads to an energy cost associated with the spatial variation of the magnetization, and this competes with the energy gain from precisely following the spatially rotating field. The result is that at intermediate values of the Zeeman field the system reduces its energy by developing an axial component of magnetization either parallel or antiparallel to the axis of the helical Zeeman field. Because the axial component of the magnetization is spatially uniform it does not contribute to the energy cost associated with the spin stiffness. By transforming to the rotating frame of the helical Zeeman field (yielding a frame with uniform spin-orbit coupling and a uniform transverse Zeeman field), one can see that this transition breaks the same symmetries as observed in the mean-field treatment of the spin-orbit coupled Bose gas~\cite{lin_Nature_2011_expt, ho_PRL_2011, ji_NatPhys_2014_expt, qizhou_NLL_PRA}. However, in the mean-field case this transition is tuned by the intensity of the Zeeman field and has been understood almost exclusively in terms of single-particle physics: for a weak Zeeman field, there are two degenerate single-particle minima related by a $Z_2$ inversion symmetry, and weak repulsive interactions favor condensation in one of these states, breaking the symmetry. At a sufficiently strong Zeeman field, the single-particle band structure changes such that there is a unique lowest-energy state in which to condense. In contrast, in the strongly interacting limit of interest to us it is more natural to understand the transition in terms of the competition between Zeeman energy and spin stiffness, the generalized rigidity associated with the interacting ferromagnetic Bose liquid. This is demonstrated in the phase diagram \fref{phases}(a), where it is clear that in contrast to the weak interaction dependence predicted by mean-field theory, the critical Zeeman field strength of the transition rapidly approaches zero at large interaction strength.

We analyze the two-component strongly interacting Bose liquid subjected to a helical Zeeman field described by the Hamiltonian density (in units with $\hbar = 1$)
\begin{align}\elabel{ham1}
\mathcal{H} = & \sum_{ss'} b^{\dagger}_s \left[ -\frac{\partial_x^2}{2m} - \mu + \frac{1}{2} \vect{\Omega}(x) \cdot \vect{\sigma} \right]_{ss'} b^{\phd}_{s'} \nonumber \\
+ & \frac{1}{2} \sum_{ss'} g_{ss'} b^{\dagger}_s b^{\phd}_s b^{\dagger}_{s'} b^{\phd}_{s'}
\end{align}
where $b = \left( b_\su,b_\sd \right)$ represents the two bosonic fields describing the physical microscopic degrees of freedom, $\vect{\sigma}$ is the vector of Pauli matrices, and $\vect{\Omega}(x) = \Omega_0 \left[ \cos( \alpha x) \vect{e}_x  - \sin( \alpha x) \vect{e}_y \right]$ is the Raman-induced helical magentic field~\cite{higbie_PRL_2002, lin_Nature_2011_expt}. From the wavevector $\alpha$ we can define a natural unit of energy $4E_R = \alpha^2/2m$, the recoil energy for the Raman laser~\cite{lin_Nature_2011_expt}. We are interested here exclusively in the effect of spin-isotropic interactions $g_{ss'} = g$.

In \secref{effective} we start from the general low-energy effective action for a spinor Bose liquid~\cite{watanabe2} and derive a low-energy effective action corresponding to \eref{ham1} that describes the Bose liquid in terms of phase (i.e., number density) and magnetization degrees of freedom. We show that magnetization fluctuations are responsible for an emergent dynamical gauge field that influences the momentum of the superfluid through spin-orbit coupling. In the presence of a helical Zeeman field we find that there is a curve of quantum critical points of the system that can be reached by tuning interaction or density; in other words, sufficiently strong interactions can disorder the $Z_2$ symmetry broken state at arbitrarily small $\Omega_0$, even though they \emph{cannot} disorder the isotropic ferromagnet in the absence of the helical field. Finally we show that spin fluctuations become locked to density fluctuations, and as a result the latter are described by an effective field theory similar to the one-dimensional Lifshitz magnet proposed originally for ferromagnetic Fermi liquids~\cite{kun_yang_fm_PRL}. The resulting critical fluctuations at the transition are qualitatively reminiscent of the non-Luttinger liquid predicted previously at the single-particle ``flat band" transition in non- or weakly-interacting spin-orbit coupled bosons~\cite{qizhou_NLL_PRA}. However, we uncover a previously unappreciated relevent interaction we find drives the system to the much less well-understood \emph{interacting} fixed point suggested in Ref.~\onlinecite{kun_yang_fm_PRL}. We also find, fortuitously, that large interaction strength decreases the length scale and increases the temperature where these critical fluctuations can be observed in experiment.
In \secref{numerics} we validate our field theory analysis with detailed density matrix renormalization group (DMRG) simulations of the microscopic model \eref{ham1}. We confirm the presence of the interaction-tuned critical point, and we find new critical exponents for the order parameter and correlation length which differ from all previously studied transitions in this model, and from the zeroth- and the first-order $\epsilon$-expansion predictions derived previously by Yang~\cite{kun_yang_fm_PRL} and by Senthil and Sachdev~\cite{sachdev_annals}. We also obtain the dynamical exponent and find it consistent with the interacting Lifshitz transition value of $z$ almost, but not exactly, $2$.

%%%%%%%%%%%%%%%%%%%%%%%%%%%%%%%%%%%%%%%%%%%%%%%%%%

\section{Effective Lagrangian approach for spinor bosons}\seclabel{effective}

The goal of this section is to analytically understand the properties of the strongly interacting Bose gas in a long-wavelength (compared to mean inter-particle spacing) helical magnetic field. Here, we restrict to this limit (analogous to Landau-Ginzburg theory) because the system is not integrable, so only universal properties such as the long wave-length limit are amenable to analytic treatment. As an added bonus, these universal properties should be relevant to other systems such as ferromagnetic Luttinger liquids.

To derive the low-energy long-wavelength properties we follow the effective field theory approach, which allows one to determine correlation functions 
from an effective Lagrangian that is determined solely by the symmetry of the system. The traditional classification of effective
field theories based on symmetry was restricted to relativistic systems, where the symmetry group includes the Lorentz group~\cite{coleman1, coleman2}.
Recent work by Watanabe and Murayama\cite{watanabe2} has extended the classification of effective Lagrangians based on symmetry breaking to non-relativistic systems, including 
systems which are in the symmetry class of the spinor Bose liquid. In Appendix~\ref{app:algebra}, we translate their results for spinor bosons, first without spin-orbit coupling, to an effective Lagrangian (up to second order in derivatives) described by a phase and a spin degree of freedom which contains only four parameters,
\begin{align}
\mathcal{L}_{\rm eff}^{C} & \;=\; -\rho_0\partial_t \phi - \frac{\rho_0}{2m}
\left[ (\partial_x \phi - a_x)^2 - v_C^{-2} (\partial_t \phi - a_t)^2 \right] \elabel{L_charge} \\
\mathcal{L}_{\rm eff}^{S} & \;=\; \rho_0 a_t - \frac{\kappa_S}{8}
\left[ ( \partial_x \vect{s})^2  - \nu_S^{-2} (\partial_t \vect{s})^2 \right] \elabel{L_spin},
\end{align}
where $\rho_0, v_C$, and $\kappa_S$ are the charge density, charge velocity, and spin-stiffness, respectively. Although $\nu_S$ has dimensions of velocity, it should not be interpreted as the spin velocity, as the spin excitations in the ferromagnetic Bose liquid disperse quadratically \cite{giamarchi_PRL_2007}.
The charge density and spin-stiffness together determine the interaction-renormalized magnon mass $m^*=\rho_0/\kappa_S$.
The degrees of freedom are a normalized vector field $\vect{s}(x,t)$ that represents the space-time texture of the spin and a phase $\phi$ whose space-time gradient $(\partial_t\phi,\partial_x \phi)$ is proportional to the fluctuations around the average density ($\rho_0$) and to the momentum density respectively. 
The fields $(a_t,a_x)$ are emergent gauge potentials satisfying the equation $\partial_t a_x-\partial_x a_t = \mathcal{E}$, where \cite{volovik}
\begin{align}\elabel{gi_flux}
&\mathcal{E}=(\partial_x \vect{s}\times\partial_t \vect{s})\cdot \vect{s}.
\end{align}
Since a gauge transformation $\vect{a}\rightarrow \vect{a}+\vect{\nabla}\Lambda$ may be off-set by a redefinition of the phase $\phi\rightarrow \phi+\Lambda$, any choice of the vector potential $(a_t,a_x)$ that has a  gauge-invariant flux given by $\mathcal{E}$ is sufficient. 

 The rather formal rigorous results (i.e. \eref{L_charge} and \eref{L_spin}) derived in Watanabe et al.~\cite{watanabe2} can also be understood intuitively on symmetry 
grounds. For example, the appearance of $\vect{s}$ and $\phi$ as the appropriate degrees of freedom is expected from the symmetry of the ferromagnetic SF ground state~\cite{eisenberg_lieb_PRL} of spinor bosons, which forms a quasi Bose condensate (hence $\phi$) and reduces the rotation symmetry from $SU(2)$ (hence $\vect{s}$) to $O(2)$. If we ignore the vector potential terms $a_{x,t}$, then \eref{L_charge} is the usual Lagrangian for the phase degree of 
freedom for a Bose condensate while \eref{L_spin} is the gradient part of the field theory description (i.e. non-linear sigma model~\cite{fradkin_book}) of a ferromagnet. The gauge potential $(a_t,a_x)$, which is essential to obtain the correct dispersion for the ferromagnetic spin waves, arises from a complication that a local rotation of the condensate about the magnetization direction $\vect{s}(x,t)$ by $\delta\phi(x,t)$ advances the phase $\phi(x,t)\rightarrow \phi(x,t)+\delta\phi(x,t)$. This is similar to how the application of a potential leads to a winding of the phase according 
to the Josephson relation. Thus, the winding of the phase in time is a combination of the applied external potential as well 
as the rotation of the condensate about the magnetization direction. To predict the effect of an external potential on the phase, 
we must therefore keep track of how much the condensate rotates along the condensate direction.

The simplest solution to this problem would be to avoid rotating the condensate around the magnetization direction $\vect{s}$. This could be 
accomplished by defining a vector $\vect{r}$ orthogonal to $\vect{s}$ (i.e. $\vect{r}\cdot\vect{s}=0$) and ensuring that $\vect{r}$ remains 
parallel as position is varied.  Since $\vect{r}\cdot\vect{s}=0$ we can  think of $\vect{r}(x,t)$ as lying on a surface (parametrized by $(x,t)$) 
which is normal (locally) to the vector field $\vect{s}(x,t)$.  The problem of choosing $\vect{r}(x,t)$ to be locally parallel at nearby points is 
exactly that of parallel transport of the vector $\vect{r}$. This turns out not to be possible because of the holonomy associated
 with the curvature of the surface defined by the magnetization $\vect{s}(x,t)$. Attempting to parallel transport $\vect{r}(x,t)$ in a small 
rectangle of size $(\delta x,\delta t)$ leads to a rotation of the vector $\vect{r}$ by an angle proportional to the Gaussian curvature $\delta\phi \simeq \vect{s}\cdot(\partial_x\vect{s}\times \partial_t \vect{s})\delta x\delta t$. This suggests that attempting to avoid rotating the condensate when position is changed 
 first along $t$ and then along $x$ leads to a net rotation about the magnetization direction compared to when the position is changed in the other order. 
This ambiguity in the net rotation leads to an ambiguity in the Berry phase that must be accounted for by a gauge potential whose curvature 
is given by \eref{gi_flux}.

\subsection{Phase diagram in a helical Zeeman field}

Given our understanding of the effective Lagrangians \eref{L_charge} and \eref{L_spin} of the fully symmetric spinor Bose gas, 
 we now investigate the effects of the helical Zeeman field. This enters into the Lagrangian as
 $\mathcal{L}_Z = \frac{\rho_0}{2}\left( 1 - \frac{\partial_t \phi}{m v_C^2} \right) \vect{\Omega}(x) \cdot \vect{s}(x,t)$. This helical Zeeman field can be unwound using a position-dependent rotation of the spin vector $\vect{s}(x,t)$ around $\vect{e}_z$. This also therefore modifies the spin gradient as $\partial_x \vect{s}\rightarrow \partial_x \vect{s}+\alpha (\vect{s} \times \vect{e}_z)$ and thus also the gauge potential as $a_x\rightarrow a_x-\alpha s_3$.
Finally the effective Lagrangian with a helical field is
\begin{widetext}
\begin{align}
\mathcal{L}_{\rm eff}^{C} & \;=\; -\rho_0\left( 1 + \frac{\Omega_0 s_1}{2mv_C^2} \right)\partial_t \phi - \frac{\rho_0}{2m}
\left[ (\partial_x \phi - a_x + \alpha s_3)^2 - v_C^{-2} (\partial_t \phi - a_t)^2 \right] \elabel{L0_charge} \\
\mathcal{L}_{\rm eff}^{S} & \;=\; \rho_0 a_t+\frac{\rho_0\Omega_0}{2} s_1 - \frac{\kappa_S}{8}
\left[ ( \partial_x \vect{s})^2 - 2 \alpha (\vect{s}\times\partial_x\vect{s})\cdot \vect{e}_{z}+ \alpha^2 (1 - s_3^2) - \nu_S^{-2} (\partial_t \vect{s})^2 \right] \elabel{L0_spin}.
\end{align}
\end{widetext}

The spin part of the Lagrangian ($\mathcal{L}_{\rm eff}^S$) is now spatially uniform, and includes an isotropic ferromagnetic exchange, a Dzyaloshinskii-Moriya term, and an easy-axis anisotropy along $\vect{e}_z$, along with a uniform Zeeman field along $\vect{e}_x$. Similar 1D spin models have been studied previously in this context, though typically in the Mott-insulating limit on a lattice, where there is no back-action on the spin from charge fluctuations~\cite{mott_xu, mott_zhao1, mott_zhao2, mott_piraud, mott_peotta, mott_pixley}.
The charge part of the Lagrangian ($\mathcal{L}_{\rm eff}^C$) now depends explicitly on spin from a dynamical vector potential $\alpha s_3$.

We now study the ground state of the system within the saddle point approximation in the limit of small $\alpha$ or large $\Omega$, where the spin aligns along $\vect{e}_x$ up to long-wavelength fluctuations. Neglecting these fluctuations, a zeroth-order saddle point approximation to the effective spin Lagrangian is
\begin{align}
\mathcal{L}_{\rm eff}^{S (0)}
= & \frac{\rho_0 \Omega_0}{2} \left( 1 - s_2^2 - s_3^2 \right)^{1/2} - \frac{\kappa_S \alpha^2}{8} (1 - s_3^2)  \nonumber \\
\approx & \left( -\frac{\kappa_S \alpha^2}{8} + \frac{\rho_0\Omega_0}{2} \right) - \frac{\rho_0\Omega_0}{4}s_2^2 \nonumber \\
& - \left( \frac{\rho_0\Omega_0}{4} - \frac{\kappa_S \alpha^2}{8} \right) s_3^2 - \frac{\rho_0 \Omega_0}{16}s_3^4 .
\end{align}
This is essentially the (real time) Landau-Ginzburg action for an Ising transition, favoring a nonzero $s_3$ when the applied helical field satisfies $\Omega_0 < \alpha^2 / 2m^*$, whether tuned by field strength, pitch angle, density, or effective spin stiffness. The magnon mass $m^*$ (or correspondingly, the spin stiffness) can be related to the dimensionless interaction strength $\gamma$ as $m^*=m/f(\gamma)$ for a known function $f$ \cite{Fuchs_PRL_2005}, such that the ferromagnetic phase in \fref{phases}(a) is given by the condition
\begin{align}
&\zeta \equiv \frac{\Omega_0}{4 E_R} < f(\gamma),
\end{align}
which recovers the mean-field result in the limit $m^* = m$.

The collective excitations of these phases are determined by the derivatives in \eref{L0_charge} and \eref{L0_spin}. To understand these excitations we fix a gauge for the vector potential using the Wess-Zumino method \cite{fradkin_book} of extending the spin-texture field $\vect{s}(x,t)$ into an extra fictitious dimension parametrized by $\lambda \in [0,1]$, so that the field $\vect{s}(x,t,\lambda=1)=\vect{s}(x,t)$ is the microscopic spin field and $\vect{s}(x,t,\lambda=0)=\vect{e}_x$. Given this extension, the vector potential is
\begin{align}
&a_j=\int d\lambda \partial_\lambda \vect{s}\cdot(\vect{s}\times \partial_j \vect{s}).\label{eqa}
\end{align}
Using the fact that for a normalized field $\vect{s}$, $\partial_\lambda \vect{s}\cdot (\partial_t \vect{s}\times \partial_x \vect{s})=0$, we can recover $\partial_t a_x-\partial_x a_t=\vect{s}\cdot (\partial_x \vect{s} \times \partial_t \vect{s})=\mathcal{E}$. We use the fact that the vector potential $a_{t,x}$ for small fluctuations $\vect{s}=\vect{e}_x+\lambda \delta\vect{s}$ simplifies from the full Wess-Zumino form (Eq.~\ref{eqa}) to $a_{j}\approx -\frac{1}{2} \vect{e}_x \cdot(\delta \vect{s}\times \partial_{j}\delta \vect{s})$. The Zeeman field gaps out the magnon modes, however the mass of the $s_3$ term vanishes near the transition and one can integrate out the massive $s_2$ degree of freedom (with mass  $\rho_0\Omega_0/4$). The dynamical term $\rho_0 s_2\partial_t s_3$ leads to the massive phase with $s_2 \simeq \partial_t s_3$. After these manipulations, we obtain a simplified effective
 Lagrangian (exact as usual at sufficiently low energies) near the phase transition,
\begin{align}
\mathcal{L}_{\rm eff}^{C} & \;=\; -\frac{\rho_0}{2m}
\left[ (\partial_x \phi +\alpha s_3)^2 - v_C^{-2} (\partial_t \phi)^2 \right] \elabel{L1_charge} \\
\mathcal{L}_{\rm eff}^{S} & \;=\;  -\frac{\kappa_S}{8}
\left[ ( \partial_x s_3)^2 - v_S^{-2} (\partial_t s_3)^2 \right] \nonumber\\
&-\left( \frac{\rho_0\Omega_0}{4} - \frac{\kappa_S \alpha^2}{8} \right) s_3^2 - \frac{\rho_0 \Omega_0}{16}s_3^4\elabel{L1_spin},
\end{align}
where the spin velocity $v_S^{-2}=\nu_S^{-2}+2\rho_0/\kappa_S\Omega_0=\nu_S^{-2}+2m^*/\Omega_0$. The term from \eref{L0_charge} that was linear in $\partial_t \phi$ has been eliminated for convenience by a constant shift $\phi(t) \rightarrow \phi(t) - A t$. We can now interpret the effective model as an Ising field theory gauge-coupled to a scalar boson. The remaining discrete symmetry of our model, however, is crucially not the usual Ising $s_3 \rightarrow -s_3$, instead it is $(s_3, \partial_x) \rightarrow (-s_3, -\partial_x)$ because of the coupling term.

\subsection{Quantum critical point}

While the phase diagram clearly suggests a $Z_2$-breaking transition where the magnetization $s_3$ orders, the itinerant nature of the magnet is expected to modify the critical properties of the transition. To understand the critical properties, we shift $s_3$ as  $s_3 \rightarrow s_3 - \partial_x\phi/\alpha$ (reflecting the symmetry above) and notice that this shifted field is gapped in the ordered phase, and so the charge and spin are locked as $s_3 = \alpha^{-1}\partial_x\phi$.
Substituting in this locking (and scaling length by $\alpha^{-1}$ and time by $(v_C \alpha)^{-1}$), the effective Lagrangian for $\phi$ alone is 
\begin{align}
\frac{\mathcal{L}}{\rho_0 E_R} \; = \; & (\partial_t\phi)^2 - f (\partial_x^2 \phi)^2
+ f \left( \frac{v_C}{v_S} \right)^{2} (\partial_t \partial_x \phi)^2 \nonumber \\
& - \left(\zeta - f \right) (\partial_x\phi)^2 - \frac{\zeta}{4}(\partial_x \phi)^4
\elabel{Lphi}
\end{align}
The third term scales to zero in the long-wavelength limit compared to the first term, so in the long-wavelength limit near the critical point we neglect it in writing down the semiclassical equations of motion
\begin{align}
\partial_t^2 \phi - \left( \zeta - f + \frac{3}{2} \zeta (\partial_x \phi)^2 \right) \partial_x^2 \phi + f \partial_x^4 \phi = 0 .
\elabel{eom}
\end{align}
We solve this numerically at the critical point and plot the resulting magnetization dynamics in \fref{phases}(b). The simulation domain is a line of length $80/\alpha$ with periodic boundary conditions, and our initial condition is a gaussian magnetization excess of height 0.02 and width $\alpha^{-1}$. We observe that the initial defect spreads with an envelope $x \propto t^{1/z}$ with $z \simeq 2$.

Finally, we estimate the length and temperature scales where we expect this treatment to be valid. The length-scale cutoff for this phase-only action near the critical point comes from the Ising model, so both $x,t$ must be larger than the inverse-gap in the spin sector, i.e. $\Lambda^{-1} \sim \alpha^{-1}\sqrt{m \kappa_S / \rho_0} = \alpha^{-1} \sqrt{f}$. Since $f$ decreases with increasing interaction, the length and inverse-temperature scales required to observe critical behavior correspondingly decrease, a boon for experimental realization.

As suggested by the estimate $z \simeq 2$ for the dynamical critical exponent, this low energy effective model we obtained is actually a ``Lifshitz" field theory and, including the relevent $(\partial_x \phi)^4$ interaction, is identical to the effective action derived by Yang for itinerant Fermi liquid ferromagnets~\cite{kun_yang_fm_PRL}. The upper critical dimension of this interaction is $d=2$, and the low-order epsilon expansion ($\epsilon = 2-d$) predictions for critical exponents~\cite{kun_yang_fm_PRL, sachdev_annals} are reproduced in \tref{eps_exp}, along with a summary of our numerical estimates presented in the following section. It was also established in previous analysis that this novel critical point has a temperature dependence for the correlation length in the quantum critical regime of $\xi \propto T^{-2}$ and spin susceptibility $\chi \propto T^{-1}$. Since the spin and density degrees of freedom are locked, the density susceptibility and correlation length also obey non-Luttinger liquid scaling at nonzero temperature~\cite{qizhou_NLL_PRA}.

\begin{table}
\caption{\tlabel{eps_exp}Scaling exponents for \eref{Lphi} near the critical point from epsilon expansion~\cite{kun_yang_fm_PRL, sachdev_annals} compared to our numerical results.}
\begin{ruledtabular}
\begin{tabular}{cccc}
$\nu$ & $\beta$ & $\eta$ & $z$ \\
$\frac{1}{2}\left( 1 + \frac{\epsilon}{6}\right)$ & $\frac{1}{2}\left( 1 - \frac{\epsilon}{3}\right)$ & $\frac{4 \epsilon^2}{225}$ & $2- \frac{\eta}{2}$ \\
$\sim \frac{1}{3}$ & $\sim \frac{1}{6}$ & $0$-$0.2$ & $1.9$-$2$ \\
\end{tabular}
\end{ruledtabular}
\end{table}

%%%%%%%%%%%%%%%%%%%%%%%%%%%%%%%%%%%%%%%%%%%%%%%%%%

\section{Numerical results}\seclabel{numerics}

We next verify and extend the conclusions of the last section by direct numerical simulation of the ground state properties of the Hamiltonian in \eref{ham1} using DMRG~\cite{itensor}. Specifically, we show that the continuous quantum phase transition shown in the phase diagram \fref{phases}(a) indeed can be accessed by varying the interaction strength at a density where the mean interparticle distance is longer than the pitch of the helical magnetic field.
We extract several critical exponents and substantiate our expectation of a Lifshitz-like critical point.
As a practical limit in simulating \eref{ham1}, we discretize the Hamiltonian on a lattice and restrict our Hilbert space to states with at most two bosons per site. This should not affect our results significantly since we consider low density (1/5 boson per site) and relatively strong interactions, described by an isotropic Hubbard interaction $U$. We emphasize that the lattice discretization is for numerical convenience only; we are not introducing a physical optical lattice potential. For most of the calculations we kept up to $800$ states to keep the truncation error per step around $10^{-12}$. However, when the interaction is very close to its critical value, we need to include more states (up to $2000$) to achieve similar truncation error. 

\begin{figure}[t]
\centering
	\includegraphics[width=0.48\textwidth]{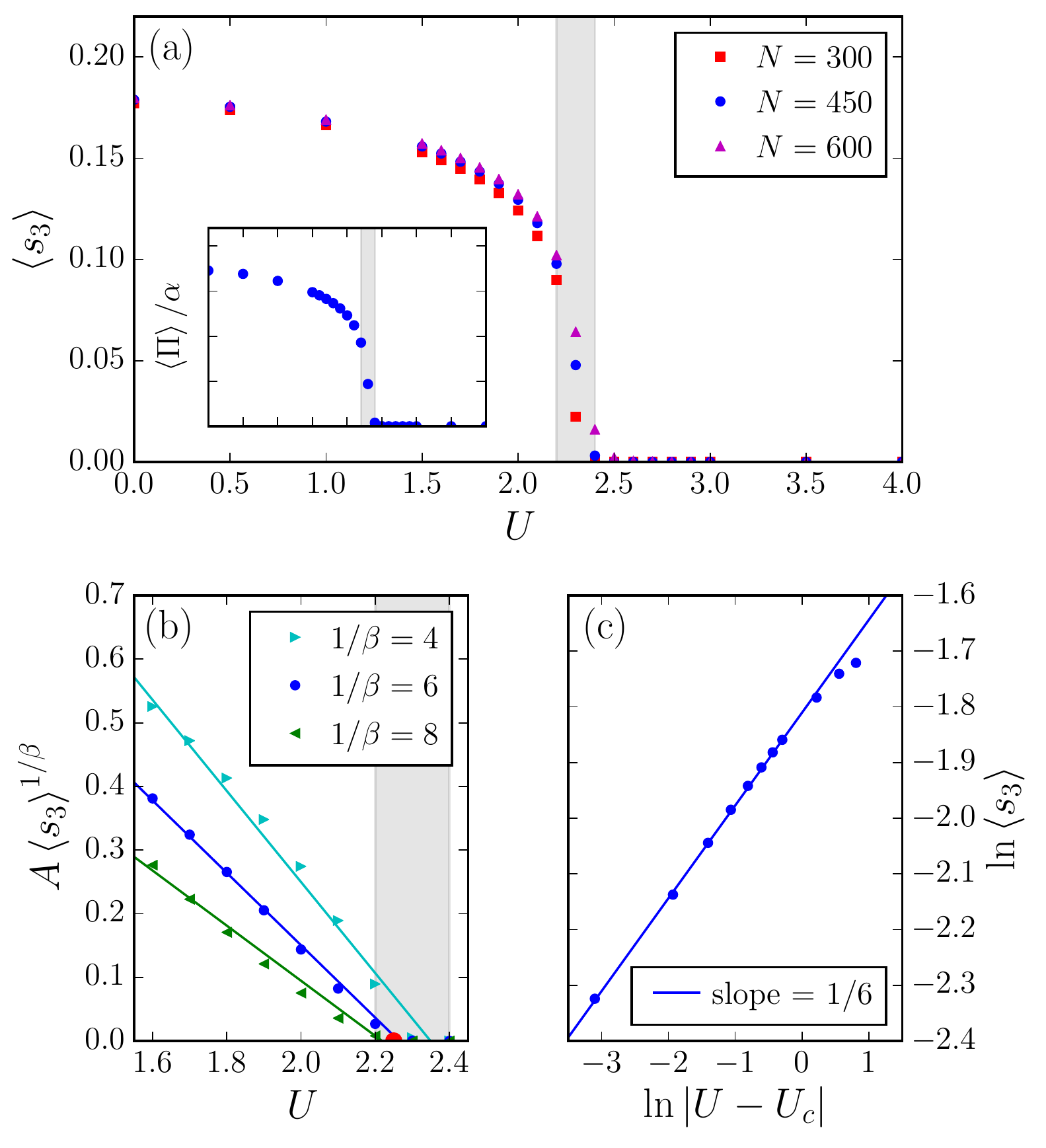}
	\caption{(a) Expectation value of magnetization as a function of interaction strength on open chains of different length. The boson density is fixed to $1/5$. An interaction driven second-order phase transition is apparent. Data in the shaded region is less-converged due to a low energy charge modulated state. (Inset) Spin-momentum locking (Eq.~(\ref{e:pisz})) is shown for $N=450$. The appropriately scaled momentum tracks the magnetization exactly. (b) For $N=450$, the critical interaction strength $U_c$ is obtained by finding the exponent that gives the best linear fit for $\langle s_3 \rangle^{1/\beta}(U)$, with $U_c$ the extrapolated intercept. We find $U_c \simeq 2.25$, indicated as a red dot. (c) Using the obtained $U_c$, we confirm the magnetization critical exponent to be $\beta \simeq 1/6$ by a linear fit on a log-log plot. }
	\label{fig:SzCorr}
\end{figure}

\subsection{Phase diagram}
The predicted $Z_2$ phase transition is identified from the magnetization expectation value $\lr{s_3}$ \cite{dmrg_sym} and correlation function $\lr{s_3(x) s_3(x')}$ on open chains while tuning the interaction strength $U$. Fig.~\ref{fig:SzCorr}(a) shows the ground state spin-density expectation value for 300, 450, and 600-site chain systems with boson density of $1/5$. The system undergoes a phase transition from a ferromagnetic ($\langle s_3 \rangle \neq 0$) to a paramagnetic ($\langle s_3 \rangle = 0$) phase with increasing interaction strength; this is the same phase transition as a horizontal cut of \fref{phases}(a). The weak dependence of the magnetization on the system size suggests that the interaction-tuned quantum phase transition is indeed continuous.

In the numerical calculations, we observe a slightly density-modulated excited state in the paramagnetic region. Such states are an unavoidable artifact of the lattice discretization we employed, and cause convergence difficulty in the calculation. We checked that running an extensive number of sweeps decreases the amplitude of the modulation and the system eventually does converge to the expected paramagnetic state with uniform density. However, in the vicinity of the critical point, where the convergence is slowest, the wave functions we obtained were not converged enough to completely eliminate the density modulation; this less-converged region gives an overestimated value of $\expect{s_3}$ and is shaded gray in Fig.~\ref{fig:SzCorr}(a) and (b).

\begin{figure}[h!]
\centering
	\includegraphics[width=0.48\textwidth]{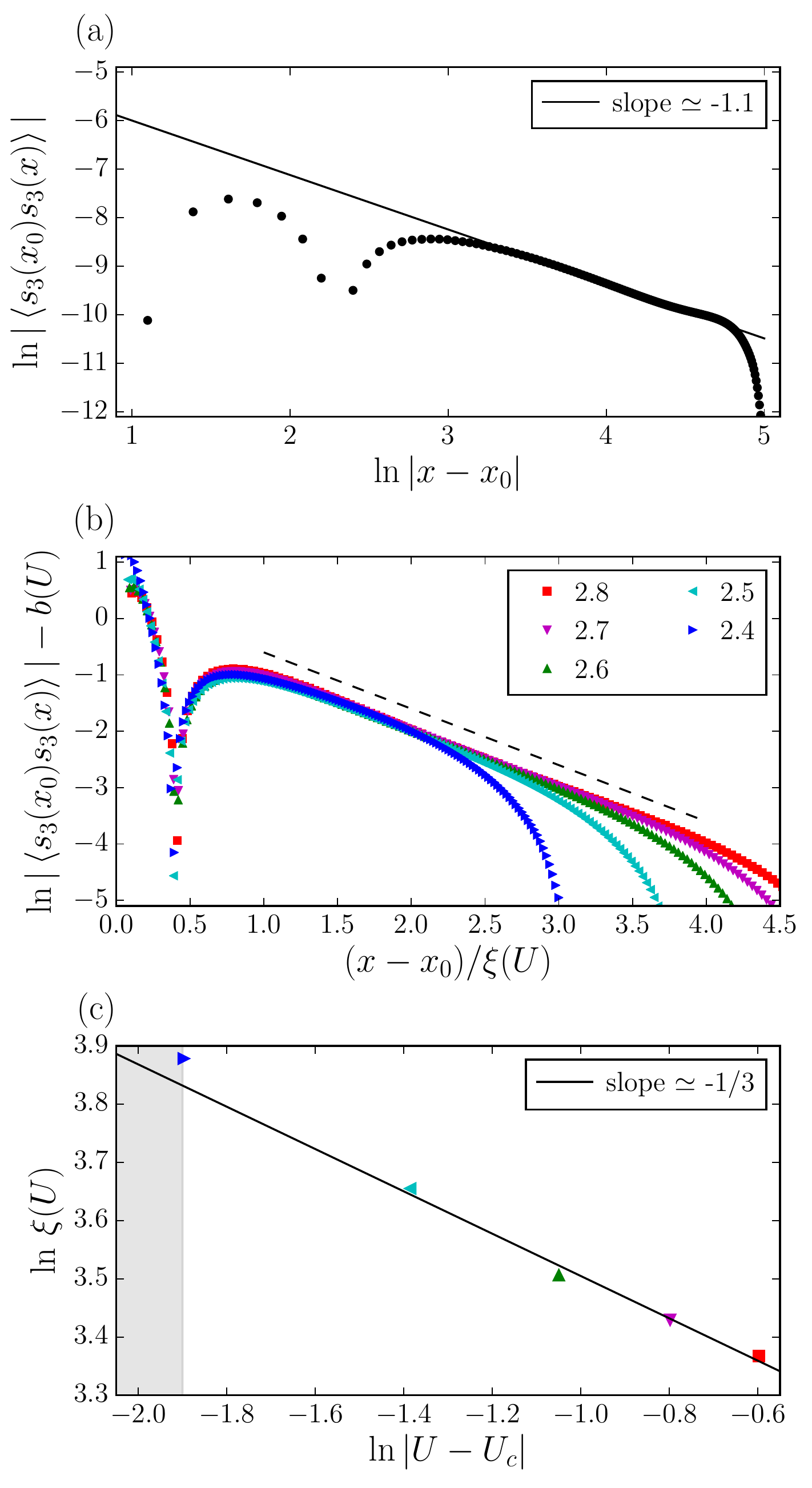}
	\caption{(a) Log-log plot of the correlation function $\langle s_3(x_0) s_3(x)\rangle$ as a function of $x$. $N=300$ sites and $U = 2.2$, which is close to the critical interaction. As per Eq.~(\ref{eq:corr}), the linear slope of the log-log plot yields $d+z-2+\eta \simeq 1.1$. (b) Data collapse of the $s_3$ correlation function. For different values of interaction (from $U = 2.4$ to $U = 2.8$) we rescale $x - x_0$ by the correlation length $\xi(U)$, to achieve data collapse. The dashed line is $e^{-(x-x_0)/\xi}$ with a small shift, as a guide to the eye. (c) Correlation length varying $U$. The solid line is a fit yielding the exponent $\nu \simeq 1/3$. The shaded region represents data collected from the less-converged wavefunctions, as in Fig.~\ref{fig:SzCorr}. }
	\label{fig:zExp}
\end{figure}

\subsection{Spin-momentum locking: signature of gauge coupling}

The role of the magnetization generated by dynamical gauge fluctuations becomes clear from the spin-momentum locking property of the ground state.  
To see the spin-momentum locking notice that the canonical momentum operator derived from the microscopic Hamiltonian is given by
\be
\Pi = i \sum_\sigma b^\dagger_\sigma \partial_x b_\sigma-(\partial_x b^\dagger_\sigma) b_\sigma 
\label{Eq:Pi}
\ee
while the number current operator derived from that same Hamiltonian is given by 
\be
j = \Pi - \alpha s_3 .
\ee
A general theorem \cite{yamamoto} rules out nonzero current density in the ground state (so $\langle j\rangle = 0$). Given this constraint, the momentum and spin-density must be related by
\be\elabel{pisz}
\lr{\Pi} = \alpha \lr{s_3} .
\ee
Thus, the gauge coupling results in all Ising-symmetry-broken ground states having finite canonical momentum. The change in momentum as one crosses the phase transition from paramagnetic to the ferromagnetic must be attributed to the effective electric field \eref{gi_flux} generated from gauge field fluctuations.

To observe the spin-momentum locking numerically, we also calculated the momentum expectation value $\lr{\Pi}$. The result, normalized by the factor $\alpha$, is plotted in the inset of Fig.~\ref{fig:SzCorr}(a). The perfect agreement of the magnetization and normalized momentum explicitly shows the spin-momentum locking, Eq.~(\ref{e:pisz}), of the system.

\subsection{Magnetization exponent}

Next we demonstrate that the transition is characterized by non-mean-field (i.e, $\epsilon = 0$) critical exponents. For this purpose, an in-depth finite-size scaling is not necessary; instead, we estimate the critical point and exponents for a sufficiently large system. 

From Fig.~\ref{fig:SzCorr}(a), one can see that finite size or boundary effects are already quite small for the $N=450$ chain. Therefore we pinpoint the critical point $U_c$ by finding the best linear fit (i.e., minimum residual) of $\expect{s_3}^{1/\beta}(U)$ over all values of $1/\beta$ (representative lines are shown in Fig.~\ref{fig:SzCorr}(b)), and then $U_c$ is the extrapolated intercept. We then confirm the magnetization critical exponent $\beta$ from a log-log plot using that value of $U_c$ in Fig.~\ref{fig:SzCorr}(c). The critical point and exponent we identified are $U_c \simeq 2.25$ and $\beta \simeq 1/6$. Note that $\beta \simeq 1/6$ is both different from both the Ising critical point ($\beta = 1/8$) and the first-order $\epsilon$-expansion result for the Lifshitz critical point in Ref.~\onlinecite{kun_yang_fm_PRL} ($\beta = 1/3$). However, we do not take $\beta \neq 1/3$ as an indication that this is not the interacting Lifshitz fixed point, but rather we suspect that the first-order $\epsilon$-expansion is unreliable.

\subsection{Dynamical critical exponent}

Now we compute the dynamical critical exponent numerically to verify whether the transition remains essentially Lifshitz (i.e. $z \simeq 2$) as estimated from the semiclassical limit in \secref{effective}.

For this purpose, we compute the equal-time connected correlation function of $s_3$. Near the critical point, i.e. when $\xi \sim N$, we expect the connected correlation function to decay as 
\begin{align}
\langle s_3(x_0) s_3(x) \rangle - \langle s_3(x_0) \rangle \langle s_3(x) \rangle\sim |x-x_0|^{-d-z+2-\eta}
\label{eq:corr}
\end{align}
The log-log plot in Fig.~\ref{fig:zExp}(a) shows that our estimated value for $d+z-2+\eta \simeq 1.1$, and we obtain the dynamical exponent to be $z \simeq 2.1 - \eta$. 

A secondary check for the value of $d+z-2+\eta$ makes use of the scaling relation
\begin{align}
	d + z - 2 + \eta = \frac{2 \beta}{\nu}
\label{eq:scaleRel}
\end{align}
The connected correlation function in Eq.~(\ref{eq:corr}) has a scaling form of $\sim F(x/\xi)$.  
We collapse the correlation function data at different values of $U$ by scaling the distance ($x$), and from the scaling values we used we are able to extract the correlation lengths at different interaction strengths.
The data collapse is shown in Fig.~\ref{fig:zExp}(b), and in Fig.~\ref{fig:zExp}(c) we plot the correlation lengths as a function of $U$. 
The correlation length decays as $\xi \sim |U-U_c|^{-\nu}$, and from the log-log plot in Fig.~\ref{fig:zExp}(c) we read off the correlation length exponent $\nu$ as $\nu \simeq 1/3$.
Plugging $\nu$ into Eq.~(\ref{eq:scaleRel}), together with $\beta \simeq 1/6$ from the previous section, again gives $d + z - 2 + \eta \simeq 1$ which is close to the result we obtained from the $s_3$ correlation function directly.

To first order in $\epsilon$-expansion, $\eta$ was predicted to be zero, but at second order the correction $z  = 2 - \eta/2$ was derived in Ref.~\onlinecite{sachdev_annals} for this critical theory. From our estimate $z \simeq 2 - \eta$ (using our estimates of $\beta$ and $\nu$) or $z \simeq 2.1 - \eta$ (using the critical correlation function directly), we therefore infer a value $\eta$ between $0$ and $0.2$, and corresponding $z$ between $2$ to $1.9$. The small but non-zero anomalous dimension, along with the other estimated critical exponents that differ from their predicted values above the upper critical dimension, indicate that this is a distinct interacting fixed point of the one-dimensional model.

%Throughout this subsection, we used the results from the $300$-site chain, which provided the best convergence of the wavefunction.

%%%%%%%%%%%%%%%%%%%%%%%%%%%%%%%%%%%%%%%%%%%%%%%%%%

\section{Discussion and conclusion}

In this work we analyzed universal properties of a strongly interacting spinor Bose gas in a helical magnetic field. We found that  
 the rigorous description of the low-energy dynamics of the spinor Bose liquid~\cite{watanabe2} 
  in terms of a scalar SF that interacts with the fluctuating magnetization as an emergent dynamical gauge field continues to apply with the addition of the helical magnetic field. The helical magnetic field is then responsible for an interaction-tuned quantum critical point, where sufficiently strong interactions can disorder an ordered Ising-like phase with a broken $Z_2$ symmetry. The effective field theory expectation of a continuous quantum phase transition was then verified with detailed DMRG simulations. Although the results are quite different, we note that as a minimal model our study also complements recent work on the 1D Ising field with an interaction (as opposed to gauge) coupling to an acoustic mode~\cite{altman_ruhman}.

Our effective field theory analysis yields a long wave-length effective Lagrangian valid near the quantum critical point identical to one proposed for one-dimensional ferromagnetic Fermi liquids~\cite{kun_yang_fm_PRL} with a Lifshitz-like dynamics (i.e. $z \simeq 2$) and similar to that that proposed for ``flat band" condensates of weakly interacting spin-orbit coupled bosons~\cite{qizhou_NLL_PRA}. In the latter case, their expectation of a collective mode with Lifshitz-like dynamics (i.e. $\omega \propto k^2$) arises straightforwardly from the underlying $k^4$ spectrum for noninteracting bosons exactly at the flat band point. In our case, strong spin-independent interactions drive the transition even for weak $\Omega_0$, far from the flat band point. Using the $\epsilon$-expansion results in Ref.~\onlinecite{kun_yang_fm_PRL} we infer that interaction-induced fluctuations modify the 
quantum critical properties from classical, mean-field estimates.
In addition to Fermi liquid ferromagnetism and the spin-orbit coupled Bose liquid studied here, \eref{ham1} can also be the starting point to describe spinless bosons in flux ladders \cite{arun, celi}, where the ``leg'' of the ladder plays the role of pseudospin and the flux is a leg-dependent hopping phase (i.e., a pseudospin-orbit coupling). This platform is subtly different because of the necessary presence of an underlying lattice, but also supports an incredibly rich landscape of quantum phases and QPTs. In future work it would be interesting to determine if the interacting fixed point we uncover here is naturally realized there as well.
More generally, Lifshitz critical points have also garnered substantial recent interest in higher spatial dimensions and in frustrated spin chains~\cite{zhai_quad, zhou_quad2, radic_quad, wu_quad, balents_quad}, and similar physics could even be relevant to superconductor-helimagnet heterostructures \cite{hals_schm}, with the superconductor providing the phase field and the direction of chirality of the helimagnet providing the Ising-like field.

The continuous phase transition in a strongly interacting gapless itinerant magnet demonstrated in this work shows that the spinor Bose liquid can be used as an experimentally realistic platform to study such quantum critical points. The combination of analytic and numerical results presented here show that the critical dynamics of the interaction-tuned $Z_2$ ferromagnetic transition differ qualitatively from mean field.   
Our results for a simpler relative of itinerant quantum critical points in Fermi liquids may yield
more general insight into those problems. The strongly interacting 1D spinor boson system in a helical magnetic field that is proposed in this work is already accessible in experiments on ultracold Rb and presents the ideal venue for the study of this class of criticality in the near future. Critical properties should be easily accessible from the temperature dependence of spatially resolved 
correlations in the system. Furthemore, given the slow timescale of the dynamics, this system
(similar to the superfluid-Mott transition~\cite{greiner}) might provide an ideal platform to observe the surprising dynamics of this critical point.

J.D.S. would like to acknowledge Ashvin Vishwanath for pointing out the connection to Ref.~\onlinecite{kun_yang_fm_PRL} and also valuable discussions with Joseph Maciejko.  W.S.C. was supported by LPS-MPO-CMTC. J.L., K.W.M., Y.A. and J.D.S. by the Alfred P. Sloan foundation, the National Science Foundation NSF DMR-1555135 (CAREER) and JQI-NSF-PFC (PHY1430094).  I.B.S. acknowledges the support of AFOSRs Quantum Matter MURI, NIST, and the NSF through the PFC at the JQI.

%\subsection*{Additional information}
%The authors declare no competing interests.

\appendix

\section{Derivation of the effective Lagrangian}\label{app:algebra}

Following Refs.~\onlinecite{coleman1, coleman2} the dynamics of the local order parameter of a system at sufficiently long wave-lengths and low 
frequencies is governed by a unitary operator $U = \exp (i\pi^a T_a)$, with the local order parameter encoded in $\pi^a(x,t)$. Their results, which lead to the classification of Goldstone 
modes in relativistic systems, have recently been extended to non-relativistic systems by Watanabe and Murayama~\cite{watanabe2} for general classes of symmetry groups. Here we are interested in the low-energy, long-wavelength Lagrangian of interacting spinor bosons, possessing $SU(2)$ rotation invariance of the spinor as well as Galilean invariance. As mentioned in the main text, the ground state spontaneously breaks Galilean invariance and reduces the rotation symmetry from $SU(2)$ to $O(2)$. From the effective field theory perspective, the low-energy dynamics is completely determined by an effective Lagrangian constrained by these symmetry (and symmetry-breaking) considerations.

In this Appendix we obtain the explicit form of the effective Lagrangian from the more general group-theoretic 
form given in Ref.~\onlinecite{watanabe2}.
We start with by recalling their results for the symmetry class of the spinor Bose gas, where the effective Lagrangian in terms of Galilean covariant derivatives contains only four parameters, up to second order in derivatives, and is written as
\begin{align}
\mathcal{L}_{\rm eff}= -&a_0 \mathcal{D}_t \pi^3 + a_1 (\mathcal{D}_t \pi^3)^2 + \nonumber \\ 
&a_2 (\mathcal{D}_t \pi^\perp)^2 - a_3 (\mathcal{D}_x \pi^\perp)^2
\end{align}
For convenience, we have combined $\pi^{1,2}$ into $\pi^\perp$. The Galilean covariant derivatives (containing the atomic mass $m$) are given as
\begin{align}
\mathcal{D}_x\pi^\perp  & = \rho_t^\perp \\
\mathcal{D}_t\pi^\perp & = \rho_t^\perp - \frac{2}{m} \rho_x^3 \rho_x^\perp
\end{align}
for $\pi^{\perp}$, while for $\pi^3$ we have
\begin{align}
\mathcal{D}_x\pi^3 & = 0 \\
\mathcal{D}_t\pi^3 & = \rho_t^3 - \frac{1}{m} (\rho_x^3)^2
\end{align}
Unpacking this notation, these covariant derivatives are constructed from components of a Maurer-Cartan form
\begin{align}
\omega \equiv -i U^\dagger d U
\end{align}
so $\omega = \omega_a d\pi^a = \omega_a^b d \pi^a T_b = \omega^b T_b$. Next, the covariant derivatives contain time and space derivatives of the $\pi^a(x,t)$. So, using
\begin{align}\elabel{om_defs}
-i U^\dagger \partial_{\mu} U = \omega_a \left( \partial_{\mu} \pi^{a} \right) 
\end{align}
suggests the shorthand notation
\be
\rho_\mu^b \equiv \omega^b_a \left( \partial_{\mu} \pi^{a} \right)
\ee
for $\mu=x,t$. (In terms of the notation of \cite{watanabe2}, $\rho_t = \bar{\omega}$, $\rho_x = \vec{\omega}$.)
Substituting this into $\mathcal{L}_{\rm eff}$ and retaining only terms up to second order in derivatives yields
\begin{align}
\mathcal{L}_{\rm eff}= -&a_0 \rho_t^3 - \frac{a_0}{m} (\rho_x^3)^2 + a_1 (\rho_t^3)^2 + \nonumber \\
&a_2 (\rho_t^\perp)^2 - a_3 (\rho_x^\perp)^2
\end{align}

Our purpose now is to translate $\mathcal{L}_{\rm eff}$, via the spacetime order parameter derivatives encoded in $\rho_\mu^b$, into experimentally relevant quantities: the magnetization and phase.
We will find it useful, following \cite{watanabe2}, to decompose the transformation $U$ as into $U=U_0 e^{i\pi^3 \sigma_0}$ where $U_0$ is now a pure $SU(2)$ rotation. ($U$ need not include the Galilean generators at this point, which were used in \cite{watanabe2} to produce the Galilean covariant derivatives). In turn, we also find it useful to express $U_0$ in terms of Euler angles $(\alpha,\beta,\gamma)$,
\begin{align}
&U_0=e^{i\alpha\sigma_3}e^{i\beta\sigma_2}e^{i\gamma\sigma_3}.
\end{align}
However, $\pi^3$ appears explicitly as the phase of $U$ and must therefore be viewed as a function $\pi^3 \equiv \pi^3(\alpha,\beta,\gamma)$.
The $\rho$ can now be written as
\begin{align}
&\rho_\mu = \Omega_\mu + \partial_\mu \pi^3 \sigma_0,
\end{align}
where 
\begin{align}
\Omega_\mu = & -i U_0^\dagger \partial_\mu U_0 \nonumber\\
 = & \; \sigma_3 (\partial_\mu \gamma + \cos{\beta}\partial_\mu \alpha) \nonumber\\
& + (\cos{\gamma} \sigma_2 + i\sin{\gamma} \sigma_1)(\partial_\mu \beta+\sin{\beta}\partial_\mu \alpha) \nonumber \\
\equiv & \; \Omega_{\mu}^b \sigma_b
\end{align}

Next, the magnetization of the spinor gas in the absence of the field points along $z$ so that $s_j\propto \delta_{j3}$. 
Any quantity which transforms like a vector and takes the value $(s_1,s_2,s_3)=(0,0,1)$  for the ferromagnetic Bose gas 
must then be proportional to the magnetization. Following this argument, the magnetization can be taken as
\begin{align}
&s_j = \Tr[U_0^\dagger \sigma_j U_0 \sigma_3]
\end{align}
This is consistent with an intuitive picture where the state of the uniform ferromagnet is taken to be $\ket{\Psi_0}$, and we are interested in the magnetization of a rotated state $U_0 \ket{\Psi_0}$,
\begin{align}
&s_j = \bra{\Psi_0} U_0^{\dagger} \sigma_j U_0 \ket{\Psi_0} = \Tr[U_0^\dagger \sigma_j U_0 \sigma_3]
\end{align}
where the trace including $\sigma_3$ reflects that only the local $z$-component of $U_0^{\dagger} \sigma_j U_0$ has non-vanishing expectation in the fully-polarized state $\ket{\Psi_0}$, as $\Tr[\sigma_j \sigma_3] = \delta_{j3}$. In terms of the Euler angles, we can first define the orthogonal transformation $U_0^{\dagger} \sigma_j U_0 = R_{j\ell} \sigma_\ell$, and then $s_j = R_{j3}$, or $(s_1,s_2,s_3) = (\sin \alpha \sin \beta, -\cos \alpha \sin \beta, \cos \beta)$, with no dependence on $\gamma$.

We can also calculate derivatives of the magnetization $s_j$ directly,
\begin{align}
\partial s_j &= \Tr[(\partial U^\dagger)\sigma_j U \sigma_3] + \Tr[U^\dagger\sigma_j (\partial U) \sigma_3] \nonumber \\
&= 2 \Tr[ U^\dagger\partial U \sigma_3 U^\dagger\sigma_j U]\\
&= 2 i R_{j \ell} \Omega^b \Tr[ \sigma_b \sigma_3 \sigma_\ell ]
\end{align}
Now, using the identity $\Tr[\sigma_b \sigma_3 \sigma_\ell]= 2i\epsilon_{b3\ell}$ and that $\rho^{\perp} = \Omega^{\perp}$ we get 
\begin{align}
\partial s_j \propto [R( \vect{e}_z \times \vect{\rho})]_j
\end{align} 
Since $R$ is an orthogonal matrix that preserves inner products,
\begin{align}
(\partial \vect{s})^2 \propto ( \vect{e}_z \times \vect{\rho} )^2 = (\rho^\perp)^2
\end{align}

The remaining terms of $\mathcal{L}_{\rm eff}$ come from the scalar field $\rho^3$. From the above, this is explicitly
\begin{align}
\rho_\mu^3 & = \partial_\mu ( \pi^3 + \gamma ) - \cos \beta \partial_\mu \alpha \nonumber \\
& \equiv \partial_\mu \phi - a_\mu
\end{align}
with the vector potential $a_\mu$ introduced to allow for the possibility of a curvature,
i.e. $\partial_t \rho_x^3-\partial_x \rho_t^3 \neq 0$.
To determine $a_\mu$, we explicitly compute
\begin{align}
\mathcal{E} & = \partial_t \rho_x^3-\partial_x \rho_t^3 \nonumber\\
& = \sin{\beta}(\partial_x\beta\partial_t\alpha-\partial_t\alpha\partial_x\beta) \nonumber \\
& = (\partial_t \vect{s}\times\partial_x \vect{s})\cdot \vect{s}
\end{align}
Therefore the vector potential $a_\mu$, which satisfies $\mathcal{E}=\partial_t a_x-\partial_x a_t$ can be chosen to depend only on $\vect{s}$
and is therefore independent of $\gamma$. That is, we can choose $\phi(\alpha,\beta,\gamma)$ as our third variable,
with the only subtlety being that $a_\mu$ must be given by the Wess-Zumino expression (i.e. Eq.~\ref{eqa}).

Substituting the various $\rho$, we obtain the final form of the low-energy long-wavelength effective Lagrangian
\begin{align}
\mathcal{L}_{\rm eff}^{C} & \;=\; -\rho_0\partial_t \phi - \frac{\rho_0}{2m}
\left[ (\partial_x \phi - a_x)^2 - v_C^{-2} (\partial_t \phi - a_t)^2 \right] \\
\mathcal{L}_{\rm eff}^{S} & \;=\; \rho_0 a_t - \frac{\kappa_S}{8}
\left[ ( \partial_x \vect{s})^2  - \nu_S^{-2} (\partial_t \vect{s})^2 \right],
\end{align}
The four previously-unassigned parameters are now given their physical significance: $\mathcal{L}_{\rm eff}^{C}$ is the real time Lagrangian of an acoustic mode with sound velocity $v_C$ in a liquid with average density $\rho_0$. $\mathcal{L}_{\rm eff}^{S}$ is the real time NL$\sigma$M Lagrangian for a spin-1/2 ferromagnet with spin-stiffness $\kappa_S$. $\nu_S$ has velocity dimensions reflecting the spacetime anisotropy, although the spin excitation spectrum is quadratic.

\section{Finite size effects in the numerics}\label{app:fss}

\begin{figure}[h]
\centering
	\includegraphics[width=0.48\textwidth]{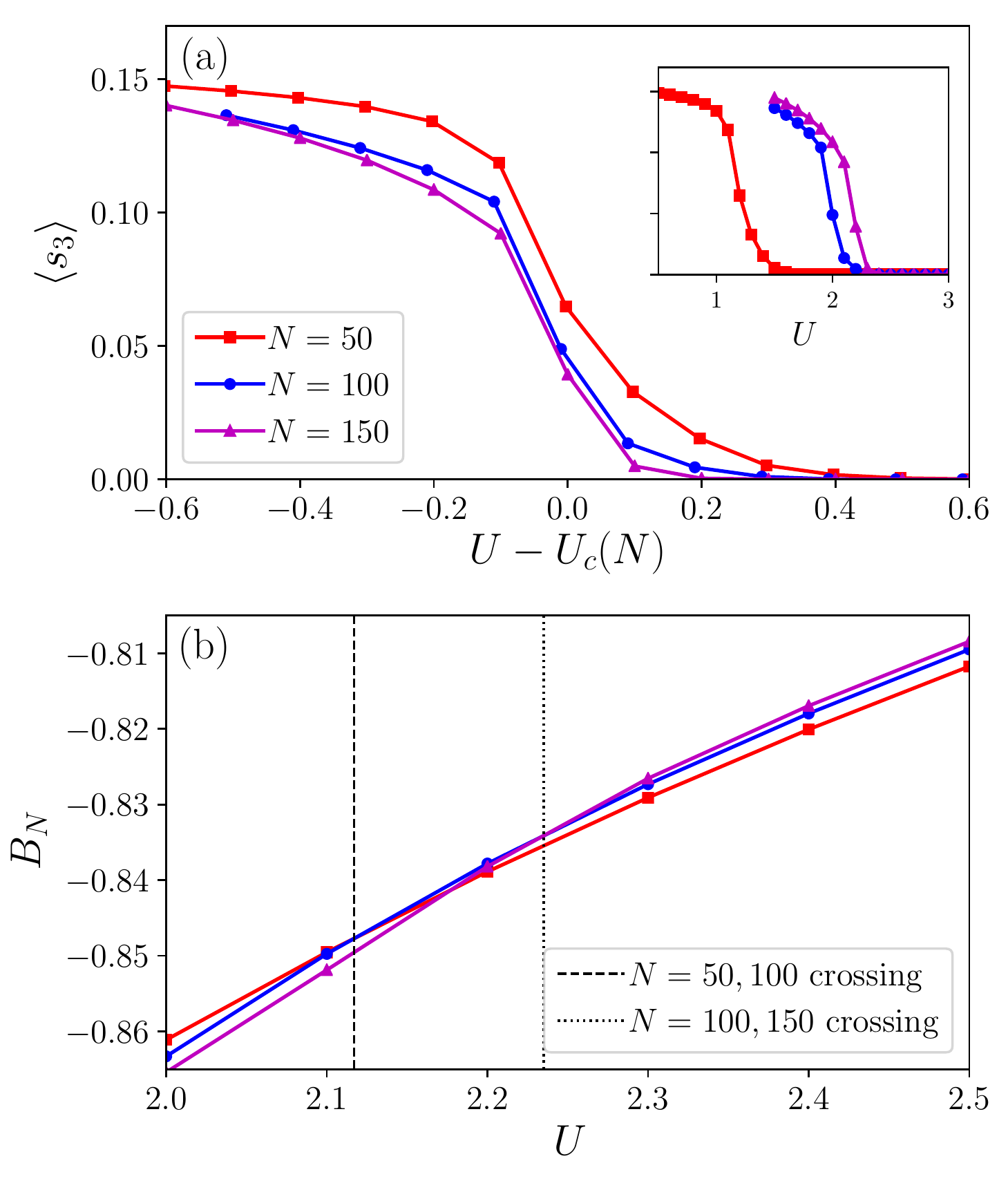}
	\caption{(a) Order parameter vs. scaled interaction, demonstrating the evolution from crossover to sharp transition with increasing system size. The axis is scaled by an estimate of the apparent critical point for comparison. The inset shows the unscaled data and the substantial finite size effect on the location of the apparent critical point. (b) Binder cumulant. The horizontal lines mark where curves for successive sizes intersect. The intersection of the $N=100$ and $150$ curves near $U \simeq 2.24$ is consistent with our infinite size estimate.}
	\label{fig:fs_figure}
\end{figure}

In \secref{numerics} we presented data for a finite length system ($N=300$) and stated that there were no appreciable finite size effects at our precision. In this Appendix we provide additional numerical results for much shorter lengths ($N=50,100,150$), where the behavior is more like a crossover. These are shown in Fig.~\ref{fig:fs_figure}. The inset of panel (a) shows $\lr{s_3}$ as a function of $U$ for each length. We can see immediately that there is a substantial finite size correction to the location of the critical point. We subtract an estimated $U_c(N)$ from each curve to show, in panel (a), the sharpening of the transition and convergence with increasing system size. In panel (b) we plot the Binder cumulant $B_N \equiv 1 - \frac{1}{3} \frac{\lr{s_3^4}}{\lr{s_3^2}^2}$. Up to non-universal finite size effects, all curves should cross at the infinite size $U_c$, so the intersections of successive sizes gives an estimate of $U_c$ that converges with increasing $N$. We find the crossing of the $N=50$ and $100$ curves occurs near $U \simeq 2.12$, while the $N=100$ and $150$ intersection occurs near $U \simeq 2.24$, already quite near our ``infinite size'' prediction of $U_c$ from the sharp transition in the $N=300$ results.

%%%%%%%%%%%%%%%%%%%%%%%%%%%%%%%%%%%%%%%%%%%%%%%%%%

%\bibliographystyle{apsrev4-1}
%\bibliography{socbosons_refs}

\begin{thebibliography}{42}%
\makeatletter
\providecommand \@ifxundefined [1]{%
 \@ifx{#1\undefined}
}%
\providecommand \@ifnum [1]{%
 \ifnum #1\expandafter \@firstoftwo
 \else \expandafter \@secondoftwo
 \fi
}%
\providecommand \@ifx [1]{%
 \ifx #1\expandafter \@firstoftwo
 \else \expandafter \@secondoftwo
 \fi
}%
\providecommand \natexlab [1]{#1}%
\providecommand \enquote  [1]{``#1''}%
\providecommand \bibnamefont  [1]{#1}%
\providecommand \bibfnamefont [1]{#1}%
\providecommand \citenamefont [1]{#1}%
\providecommand \href@noop [0]{\@secondoftwo}%
\providecommand \href [0]{\begingroup \@sanitize@url \@href}%
\providecommand \@href[1]{\@@startlink{#1}\@@href}%
\providecommand \@@href[1]{\endgroup#1\@@endlink}%
\providecommand \@sanitize@url [0]{\catcode `\\12\catcode `\$12\catcode
  `\&12\catcode `\#12\catcode `\^12\catcode `\_12\catcode `\%12\relax}%
\providecommand \@@startlink[1]{}%
\providecommand \@@endlink[0]{}%
\providecommand \url  [0]{\begingroup\@sanitize@url \@url }%
\providecommand \@url [1]{\endgroup\@href {#1}{\urlprefix }}%
\providecommand \urlprefix  [0]{URL }%
\providecommand \Eprint [0]{\href }%
\providecommand \doibase [0]{http://dx.doi.org/}%
\providecommand \selectlanguage [0]{\@gobble}%
\providecommand \bibinfo  [0]{\@secondoftwo}%
\providecommand \bibfield  [0]{\@secondoftwo}%
\providecommand \translation [1]{[#1]}%
\providecommand \BibitemOpen [0]{}%
\providecommand \bibitemStop [0]{}%
\providecommand \bibitemNoStop [0]{.\EOS\space}%
\providecommand \EOS [0]{\spacefactor3000\relax}%
\providecommand \BibitemShut  [1]{\csname bibitem#1\endcsname}%
\let\auto@bib@innerbib\@empty
%</preamble>
\bibitem [{\citenamefont {Hertz}(1976)}]{hertz}%
  \BibitemOpen
  \bibfield  {author} {\bibinfo {author} {\bibfnamefont {John~A.}\ \bibnamefont
  {Hertz}},\ }\bibfield  {title} {\enquote {\bibinfo {title} {Quantum critical
  phenomena},}\ }\href {\doibase 10.1103/PhysRevB.14.1165} {\bibfield
  {journal} {\bibinfo  {journal} {Phys. Rev. B}\ }\textbf {\bibinfo {volume}
  {14}},\ \bibinfo {pages} {1165--1184} (\bibinfo {year} {1976})}\BibitemShut
  {NoStop}%
\bibitem [{\citenamefont {Millis}(1993)}]{millis}%
  \BibitemOpen
  \bibfield  {author} {\bibinfo {author} {\bibfnamefont {A.~J.}\ \bibnamefont
  {Millis}},\ }\bibfield  {title} {\enquote {\bibinfo {title} {Effect of a
  nonzero temperature on quantum critical points in itinerant fermion
  systems},}\ }\href {\doibase 10.1103/PhysRevB.48.7183} {\bibfield  {journal}
  {\bibinfo  {journal} {Phys. Rev. B}\ }\textbf {\bibinfo {volume} {48}},\
  \bibinfo {pages} {7183--7196} (\bibinfo {year} {1993})}\BibitemShut {NoStop}%
\bibitem [{\citenamefont {Sachdev}(2011)}]{sachdev_book}%
  \BibitemOpen
  \bibfield  {author} {\bibinfo {author} {\bibfnamefont {S.}~\bibnamefont
  {Sachdev}},\ }\href@noop {} {\emph {\bibinfo {title} {Quantum Phase
  Transitions}}}\ (\bibinfo  {publisher} {Cambridge University Press},\
  \bibinfo {year} {2011})\BibitemShut {NoStop}%
\bibitem [{\citenamefont {Yang}(2004)}]{kun_yang_fm_PRL}%
  \BibitemOpen
  \bibfield  {author} {\bibinfo {author} {\bibfnamefont {Kun}\ \bibnamefont
  {Yang}},\ }\bibfield  {title} {\enquote {\bibinfo {title} {Ferromagnetic
  transition in one-dimensional itinerant electron systems},}\ }\href {\doibase
  10.1103/PhysRevLett.93.066401} {\bibfield  {journal} {\bibinfo  {journal}
  {Phys. Rev. Lett.}\ }\textbf {\bibinfo {volume} {93}},\ \bibinfo {pages}
  {066401} (\bibinfo {year} {2004})}\BibitemShut {NoStop}%
\bibitem [{\citenamefont {Kozii}\ \emph {et~al.}(2017)\citenamefont {Kozii},
  \citenamefont {Ruhman}, \citenamefont {Fu},\ and\ \citenamefont
  {Radzihovsky}}]{kozii_2017}%
  \BibitemOpen
  \bibfield  {author} {\bibinfo {author} {\bibfnamefont {Vladyslav}\
  \bibnamefont {Kozii}}, \bibinfo {author} {\bibfnamefont {Jonathan}\
  \bibnamefont {Ruhman}}, \bibinfo {author} {\bibfnamefont {Liang}\
  \bibnamefont {Fu}}, \ and\ \bibinfo {author} {\bibfnamefont {Leo}\
  \bibnamefont {Radzihovsky}},\ }\bibfield  {title} {\enquote {\bibinfo {title}
  {Ferromagnetic transition in a one-dimensional spin-orbit-coupled metal and
  its mapping to a critical point in smectic liquid crystals},}\ }\href
  {\doibase 10.1103/PhysRevB.96.094419} {\bibfield  {journal} {\bibinfo
  {journal} {Phys. Rev. B}\ }\textbf {\bibinfo {volume} {96}},\ \bibinfo
  {pages} {094419} (\bibinfo {year} {2017})}\BibitemShut {NoStop}%
\bibitem [{\citenamefont {Eisenberg}\ and\ \citenamefont
  {Lieb}(2002)}]{eisenberg_lieb_PRL}%
  \BibitemOpen
  \bibfield  {author} {\bibinfo {author} {\bibfnamefont {Eli}\ \bibnamefont
  {Eisenberg}}\ and\ \bibinfo {author} {\bibfnamefont {Elliott~H.}\
  \bibnamefont {Lieb}},\ }\bibfield  {title} {\enquote {\bibinfo {title}
  {Polarization of interacting bosons with spin},}\ }\href {\doibase
  10.1103/PhysRevLett.89.220403} {\bibfield  {journal} {\bibinfo  {journal}
  {Phys. Rev. Lett.}\ }\textbf {\bibinfo {volume} {89}},\ \bibinfo {pages}
  {220403} (\bibinfo {year} {2002})}\BibitemShut {NoStop}%
\bibitem [{\citenamefont {Higbie}\ and\ \citenamefont
  {Stamper-Kurn}(2002)}]{higbie_PRL_2002}%
  \BibitemOpen
  \bibfield  {author} {\bibinfo {author} {\bibfnamefont {J.}~\bibnamefont
  {Higbie}}\ and\ \bibinfo {author} {\bibfnamefont {D.~M.}\ \bibnamefont
  {Stamper-Kurn}},\ }\bibfield  {title} {\enquote {\bibinfo {title}
  {Periodically dressed bose-einstein condensate: A superfluid with an
  anisotropic and variable critical velocity},}\ }\href {\doibase
  10.1103/PhysRevLett.88.090401} {\bibfield  {journal} {\bibinfo  {journal}
  {Phys. Rev. Lett.}\ }\textbf {\bibinfo {volume} {88}},\ \bibinfo {pages}
  {090401} (\bibinfo {year} {2002})}\BibitemShut {NoStop}%
\bibitem [{\citenamefont {{Lin}}\ \emph {et~al.}(2011)\citenamefont {{Lin}},
  \citenamefont {{Jim{\'e}nez-Garc{\'{\i}}a}},\ and\ \citenamefont
  {{Spielman}}}]{lin_Nature_2011_expt}%
  \BibitemOpen
  \bibfield  {author} {\bibinfo {author} {\bibfnamefont {Y.-J.}\ \bibnamefont
  {{Lin}}}, \bibinfo {author} {\bibfnamefont {K.}~\bibnamefont
  {{Jim{\'e}nez-Garc{\'{\i}}a}}}, \ and\ \bibinfo {author} {\bibfnamefont
  {I.~B.}\ \bibnamefont {{Spielman}}},\ }\bibfield  {title} {\enquote {\bibinfo
  {title} {{Spin-orbit-coupled Bose-Einstein condensates}},}\ }\href {\doibase
  10.1038/nature09887} {\bibfield  {journal} {\bibinfo  {journal} {Nature}\
  }\textbf {\bibinfo {volume} {471}},\ \bibinfo {pages} {83--86} (\bibinfo
  {year} {2011})}\BibitemShut {NoStop}%
\bibitem [{\citenamefont {Ho}\ and\ \citenamefont {Zhang}(2011)}]{ho_PRL_2011}%
  \BibitemOpen
  \bibfield  {author} {\bibinfo {author} {\bibfnamefont {Tin-Lun}\ \bibnamefont
  {Ho}}\ and\ \bibinfo {author} {\bibfnamefont {Shizhong}\ \bibnamefont
  {Zhang}},\ }\bibfield  {title} {\enquote {\bibinfo {title} {Bose-einstein
  condensates with spin-orbit interaction},}\ }\href {\doibase
  10.1103/PhysRevLett.107.150403} {\bibfield  {journal} {\bibinfo  {journal}
  {Phys. Rev. Lett.}\ }\textbf {\bibinfo {volume} {107}},\ \bibinfo {pages}
  {150403} (\bibinfo {year} {2011})}\BibitemShut {NoStop}%
\bibitem [{\citenamefont {Li}\ \emph {et~al.}(2012)\citenamefont {Li},
  \citenamefont {Pitaevskii},\ and\ \citenamefont {Stringari}}]{li_PRL_2012}%
  \BibitemOpen
  \bibfield  {author} {\bibinfo {author} {\bibfnamefont {Yun}\ \bibnamefont
  {Li}}, \bibinfo {author} {\bibfnamefont {Lev~P.}\ \bibnamefont {Pitaevskii}},
  \ and\ \bibinfo {author} {\bibfnamefont {Sandro}\ \bibnamefont {Stringari}},\
  }\bibfield  {title} {\enquote {\bibinfo {title} {Quantum tricriticality and
  phase transitions in spin-orbit coupled bose-einstein condensates},}\ }\href
  {\doibase 10.1103/PhysRevLett.108.225301} {\bibfield  {journal} {\bibinfo
  {journal} {Phys. Rev. Lett.}\ }\textbf {\bibinfo {volume} {108}},\ \bibinfo
  {pages} {225301} (\bibinfo {year} {2012})}\BibitemShut {NoStop}%
\bibitem [{\citenamefont {Zhang}\ \emph {et~al.}(2012)\citenamefont {Zhang},
  \citenamefont {Ji}, \citenamefont {Chen}, \citenamefont {Zhang},
  \citenamefont {Du}, \citenamefont {Yan}, \citenamefont {Pan}, \citenamefont
  {Zhao}, \citenamefont {Deng}, \citenamefont {Zhai}, \citenamefont {Chen},\
  and\ \citenamefont {Pan}}]{zhang_PRL_2012_expt}%
  \BibitemOpen
  \bibfield  {author} {\bibinfo {author} {\bibfnamefont {Jin-Yi}\ \bibnamefont
  {Zhang}}, \bibinfo {author} {\bibfnamefont {Si-Cong}\ \bibnamefont {Ji}},
  \bibinfo {author} {\bibfnamefont {Zhu}\ \bibnamefont {Chen}}, \bibinfo
  {author} {\bibfnamefont {Long}\ \bibnamefont {Zhang}}, \bibinfo {author}
  {\bibfnamefont {Zhi-Dong}\ \bibnamefont {Du}}, \bibinfo {author}
  {\bibfnamefont {Bo}~\bibnamefont {Yan}}, \bibinfo {author} {\bibfnamefont
  {Ge-Sheng}\ \bibnamefont {Pan}}, \bibinfo {author} {\bibfnamefont
  {Bo}~\bibnamefont {Zhao}}, \bibinfo {author} {\bibfnamefont {You-Jin}\
  \bibnamefont {Deng}}, \bibinfo {author} {\bibfnamefont {Hui}\ \bibnamefont
  {Zhai}}, \bibinfo {author} {\bibfnamefont {Shuai}\ \bibnamefont {Chen}}, \
  and\ \bibinfo {author} {\bibfnamefont {Jian-Wei}\ \bibnamefont {Pan}},\
  }\bibfield  {title} {\enquote {\bibinfo {title} {Collective dipole
  oscillations of a spin-orbit coupled bose-einstein condensate},}\ }\href
  {\doibase 10.1103/PhysRevLett.109.115301} {\bibfield  {journal} {\bibinfo
  {journal} {Phys. Rev. Lett.}\ }\textbf {\bibinfo {volume} {109}},\ \bibinfo
  {pages} {115301} (\bibinfo {year} {2012})}\BibitemShut {NoStop}%
\bibitem [{\citenamefont {{Beeler}}\ \emph {et~al.}(2013)\citenamefont
  {{Beeler}}, \citenamefont {{Williams}}, \citenamefont
  {{Jim{\'e}nez-Garc{\'{\i}}a}}, \citenamefont {{Leblanc}}, \citenamefont
  {{Perry}},\ and\ \citenamefont {{Spielman}}}]{beeler_Nature_2013_expt}%
  \BibitemOpen
  \bibfield  {author} {\bibinfo {author} {\bibfnamefont {M.~C.}\ \bibnamefont
  {{Beeler}}}, \bibinfo {author} {\bibfnamefont {R.~A.}\ \bibnamefont
  {{Williams}}}, \bibinfo {author} {\bibfnamefont {K.}~\bibnamefont
  {{Jim{\'e}nez-Garc{\'{\i}}a}}}, \bibinfo {author} {\bibfnamefont {L.~J.}\
  \bibnamefont {{Leblanc}}}, \bibinfo {author} {\bibfnamefont {A.~R.}\
  \bibnamefont {{Perry}}}, \ and\ \bibinfo {author} {\bibfnamefont {I.~B.}\
  \bibnamefont {{Spielman}}},\ }\bibfield  {title} {\enquote {\bibinfo {title}
  {{The spin Hall effect in a quantum gas}},}\ }\href {\doibase
  10.1038/nature12185} {\bibfield  {journal} {\bibinfo  {journal} {Nature}\
  }\textbf {\bibinfo {volume} {498}},\ \bibinfo {pages} {201--204} (\bibinfo
  {year} {2013})}\BibitemShut {NoStop}%
\bibitem [{\citenamefont {Barnes}\ and\ \citenamefont
  {Maekawa}(2005)}]{barnes_PRL_2005}%
  \BibitemOpen
  \bibfield  {author} {\bibinfo {author} {\bibfnamefont {S.~E.}\ \bibnamefont
  {Barnes}}\ and\ \bibinfo {author} {\bibfnamefont {S.}~\bibnamefont
  {Maekawa}},\ }\bibfield  {title} {\enquote {\bibinfo {title} {Current-spin
  coupling for ferromagnetic domain walls in fine wires},}\ }\href {\doibase
  10.1103/PhysRevLett.95.107204} {\bibfield  {journal} {\bibinfo  {journal}
  {Phys. Rev. Lett.}\ }\textbf {\bibinfo {volume} {95}},\ \bibinfo {pages}
  {107204} (\bibinfo {year} {2005})}\BibitemShut {NoStop}%
\bibitem [{\citenamefont {{Ji}}\ \emph {et~al.}(2014)\citenamefont {{Ji}},
  \citenamefont {{Zhang}}, \citenamefont {{Zhang}}, \citenamefont {{Du}},
  \citenamefont {{Zheng}}, \citenamefont {{Deng}}, \citenamefont {{Zhai}},
  \citenamefont {{Chen}},\ and\ \citenamefont {{Pan}}}]{ji_NatPhys_2014_expt}%
  \BibitemOpen
  \bibfield  {author} {\bibinfo {author} {\bibfnamefont {S.-C.}\ \bibnamefont
  {{Ji}}}, \bibinfo {author} {\bibfnamefont {J.-Y.}\ \bibnamefont {{Zhang}}},
  \bibinfo {author} {\bibfnamefont {L.}~\bibnamefont {{Zhang}}}, \bibinfo
  {author} {\bibfnamefont {Z.-D.}\ \bibnamefont {{Du}}}, \bibinfo {author}
  {\bibfnamefont {W.}~\bibnamefont {{Zheng}}}, \bibinfo {author} {\bibfnamefont
  {Y.-J.}\ \bibnamefont {{Deng}}}, \bibinfo {author} {\bibfnamefont
  {H.}~\bibnamefont {{Zhai}}}, \bibinfo {author} {\bibfnamefont
  {S.}~\bibnamefont {{Chen}}}, \ and\ \bibinfo {author} {\bibfnamefont {J.-W.}\
  \bibnamefont {{Pan}}},\ }\bibfield  {title} {\enquote {\bibinfo {title}
  {{Experimental determination of the finite-temperature phase diagram of a
  spin-orbit coupled Bose gas}},}\ }\href {\doibase 10.1038/nphys2905}
  {\bibfield  {journal} {\bibinfo  {journal} {Nature Physics}\ }\textbf
  {\bibinfo {volume} {10}},\ \bibinfo {pages} {314--320} (\bibinfo {year}
  {2014})}\BibitemShut {NoStop}%
\bibitem [{\citenamefont {Po}\ \emph {et~al.}(2014)\citenamefont {Po},
  \citenamefont {Chen},\ and\ \citenamefont {Zhou}}]{qizhou_NLL_PRA}%
  \BibitemOpen
  \bibfield  {author} {\bibinfo {author} {\bibfnamefont {Hoi~Chun}\
  \bibnamefont {Po}}, \bibinfo {author} {\bibfnamefont {Weiqiang}\ \bibnamefont
  {Chen}}, \ and\ \bibinfo {author} {\bibfnamefont {Qi}~\bibnamefont {Zhou}},\
  }\bibfield  {title} {\enquote {\bibinfo {title} {Non-luttinger quantum liquid
  of one-dimensional spin-orbit-coupled bosons},}\ }\href {\doibase
  10.1103/PhysRevA.90.011602} {\bibfield  {journal} {\bibinfo  {journal} {Phys.
  Rev. A}\ }\textbf {\bibinfo {volume} {90}},\ \bibinfo {pages} {011602}
  (\bibinfo {year} {2014})}\BibitemShut {NoStop}%
\bibitem [{\citenamefont {Watanabe}\ and\ \citenamefont
  {Murayama}(2014)}]{watanabe2}%
  \BibitemOpen
  \bibfield  {author} {\bibinfo {author} {\bibfnamefont {Haruki}\ \bibnamefont
  {Watanabe}}\ and\ \bibinfo {author} {\bibfnamefont {Hitoshi}\ \bibnamefont
  {Murayama}},\ }\bibfield  {title} {\enquote {\bibinfo {title} {Effective
  lagrangian for nonrelativistic systems},}\ }\href {\doibase
  10.1103/PhysRevX.4.031057} {\bibfield  {journal} {\bibinfo  {journal} {Phys.
  Rev. X}\ }\textbf {\bibinfo {volume} {4}},\ \bibinfo {pages} {031057}
  (\bibinfo {year} {2014})}\BibitemShut {NoStop}%
\bibitem [{\citenamefont {Sachdev}\ and\ \citenamefont
  {Senthil}(1996)}]{sachdev_annals}%
  \BibitemOpen
  \bibfield  {author} {\bibinfo {author} {\bibfnamefont {Subir}\ \bibnamefont
  {Sachdev}}\ and\ \bibinfo {author} {\bibfnamefont {T.}~\bibnamefont
  {Senthil}},\ }\bibfield  {title} {\enquote {\bibinfo {title} {Zero
  temperature phase transitions in quantum heisenberg ferromagnets},}\ }\href
  {\doibase https://doi.org/10.1006/aphy.1996.0108} {\bibfield  {journal}
  {\bibinfo  {journal} {Annals of Physics}\ }\textbf {\bibinfo {volume}
  {251}},\ \bibinfo {pages} {76 -- 122} (\bibinfo {year} {1996})}\BibitemShut
  {NoStop}%
\bibitem [{\citenamefont {Coleman}\ \emph {et~al.}(1969)\citenamefont
  {Coleman}, \citenamefont {Wess},\ and\ \citenamefont {Zumino}}]{coleman1}%
  \BibitemOpen
  \bibfield  {author} {\bibinfo {author} {\bibfnamefont {S.}~\bibnamefont
  {Coleman}}, \bibinfo {author} {\bibfnamefont {J.}~\bibnamefont {Wess}}, \
  and\ \bibinfo {author} {\bibfnamefont {Bruno}\ \bibnamefont {Zumino}},\
  }\bibfield  {title} {\enquote {\bibinfo {title} {Structure of
  phenomenological lagrangians. i},}\ }\href {\doibase
  10.1103/PhysRev.177.2239} {\bibfield  {journal} {\bibinfo  {journal} {Phys.
  Rev.}\ }\textbf {\bibinfo {volume} {177}},\ \bibinfo {pages} {2239--2247}
  (\bibinfo {year} {1969})}\BibitemShut {NoStop}%
\bibitem [{\citenamefont {Callan}\ \emph {et~al.}(1969)\citenamefont {Callan},
  \citenamefont {Coleman}, \citenamefont {Wess},\ and\ \citenamefont
  {Zumino}}]{coleman2}%
  \BibitemOpen
  \bibfield  {author} {\bibinfo {author} {\bibfnamefont {Curtis~G.}\
  \bibnamefont {Callan}}, \bibinfo {author} {\bibfnamefont {Sidney}\
  \bibnamefont {Coleman}}, \bibinfo {author} {\bibfnamefont {J.}~\bibnamefont
  {Wess}}, \ and\ \bibinfo {author} {\bibfnamefont {Bruno}\ \bibnamefont
  {Zumino}},\ }\bibfield  {title} {\enquote {\bibinfo {title} {Structure of
  phenomenological lagrangians. ii},}\ }\href {\doibase
  10.1103/PhysRev.177.2247} {\bibfield  {journal} {\bibinfo  {journal} {Phys.
  Rev.}\ }\textbf {\bibinfo {volume} {177}},\ \bibinfo {pages} {2247--2250}
  (\bibinfo {year} {1969})}\BibitemShut {NoStop}%
\bibitem [{\citenamefont {Zvonarev}\ \emph {et~al.}(2007)\citenamefont
  {Zvonarev}, \citenamefont {Cheianov},\ and\ \citenamefont
  {Giamarchi}}]{giamarchi_PRL_2007}%
  \BibitemOpen
  \bibfield  {author} {\bibinfo {author} {\bibfnamefont {M.~B.}\ \bibnamefont
  {Zvonarev}}, \bibinfo {author} {\bibfnamefont {V.~V.}\ \bibnamefont
  {Cheianov}}, \ and\ \bibinfo {author} {\bibfnamefont {T.}~\bibnamefont
  {Giamarchi}},\ }\bibfield  {title} {\enquote {\bibinfo {title} {Spin dynamics
  in a one-dimensional ferromagnetic bose gas},}\ }\href {\doibase
  10.1103/PhysRevLett.99.240404} {\bibfield  {journal} {\bibinfo  {journal}
  {Phys. Rev. Lett.}\ }\textbf {\bibinfo {volume} {99}},\ \bibinfo {pages}
  {240404} (\bibinfo {year} {2007})}\BibitemShut {NoStop}%
\bibitem [{\citenamefont {Volovik}(1987)}]{volovik}%
  \BibitemOpen
  \bibfield  {author} {\bibinfo {author} {\bibfnamefont {G~E}\ \bibnamefont
  {Volovik}},\ }\bibfield  {title} {\enquote {\bibinfo {title} {Linear momentum
  in ferromagnets},}\ }\href {http://stacks.iop.org/0022-3719/20/i=7/a=003}
  {\bibfield  {journal} {\bibinfo  {journal} {Journal of Physics C: Solid State
  Physics}\ }\textbf {\bibinfo {volume} {20}},\ \bibinfo {pages} {L83}
  (\bibinfo {year} {1987})}\BibitemShut {NoStop}%
\bibitem [{\citenamefont {Fradkin}(2013)}]{fradkin_book}%
  \BibitemOpen
  \bibfield  {author} {\bibinfo {author} {\bibfnamefont {E.}~\bibnamefont
  {Fradkin}},\ }\href@noop {} {\emph {\bibinfo {title} {Field Theories of
  Condensed Matter Physics}}}\ (\bibinfo  {publisher} {Cambridge University
  Press},\ \bibinfo {year} {2013})\BibitemShut {NoStop}%
\bibitem [{\citenamefont {Xu}\ \emph {et~al.}(2014)\citenamefont {Xu},
  \citenamefont {Cole},\ and\ \citenamefont {Zhang}}]{mott_xu}%
  \BibitemOpen
  \bibfield  {author} {\bibinfo {author} {\bibfnamefont {Zhihao}\ \bibnamefont
  {Xu}}, \bibinfo {author} {\bibfnamefont {William~S.}\ \bibnamefont {Cole}}, \
  and\ \bibinfo {author} {\bibfnamefont {Shizhong}\ \bibnamefont {Zhang}},\
  }\bibfield  {title} {\enquote {\bibinfo {title} {Mott-superfluid transition
  for spin-orbit-coupled bosons in one-dimensional optical lattices},}\ }\href
  {\doibase 10.1103/PhysRevA.89.051604} {\bibfield  {journal} {\bibinfo
  {journal} {Phys. Rev. A}\ }\textbf {\bibinfo {volume} {89}},\ \bibinfo
  {pages} {051604} (\bibinfo {year} {2014})}\BibitemShut {NoStop}%
\bibitem [{\citenamefont {Zhao}\ \emph
  {et~al.}(2014{\natexlab{a}})\citenamefont {Zhao}, \citenamefont {Hu},
  \citenamefont {Chang}, \citenamefont {Zhang},\ and\ \citenamefont
  {Wang}}]{mott_zhao1}%
  \BibitemOpen
  \bibfield  {author} {\bibinfo {author} {\bibfnamefont {Jize}\ \bibnamefont
  {Zhao}}, \bibinfo {author} {\bibfnamefont {Shijie}\ \bibnamefont {Hu}},
  \bibinfo {author} {\bibfnamefont {Jun}\ \bibnamefont {Chang}}, \bibinfo
  {author} {\bibfnamefont {Ping}\ \bibnamefont {Zhang}}, \ and\ \bibinfo
  {author} {\bibfnamefont {Xiaoqun}\ \bibnamefont {Wang}},\ }\bibfield  {title}
  {\enquote {\bibinfo {title} {Ferromagnetism in a two-component bose-hubbard
  model with synthetic spin-orbit coupling},}\ }\href {\doibase
  10.1103/PhysRevA.89.043611} {\bibfield  {journal} {\bibinfo  {journal} {Phys.
  Rev. A}\ }\textbf {\bibinfo {volume} {89}},\ \bibinfo {pages} {043611}
  (\bibinfo {year} {2014}{\natexlab{a}})}\BibitemShut {NoStop}%
\bibitem [{\citenamefont {Zhao}\ \emph
  {et~al.}(2014{\natexlab{b}})\citenamefont {Zhao}, \citenamefont {Hu},
  \citenamefont {Chang}, \citenamefont {Zheng}, \citenamefont {Zhang},\ and\
  \citenamefont {Wang}}]{mott_zhao2}%
  \BibitemOpen
  \bibfield  {author} {\bibinfo {author} {\bibfnamefont {Jize}\ \bibnamefont
  {Zhao}}, \bibinfo {author} {\bibfnamefont {Shijie}\ \bibnamefont {Hu}},
  \bibinfo {author} {\bibfnamefont {Jun}\ \bibnamefont {Chang}}, \bibinfo
  {author} {\bibfnamefont {Fawei}\ \bibnamefont {Zheng}}, \bibinfo {author}
  {\bibfnamefont {Ping}\ \bibnamefont {Zhang}}, \ and\ \bibinfo {author}
  {\bibfnamefont {Xiaoqun}\ \bibnamefont {Wang}},\ }\bibfield  {title}
  {\enquote {\bibinfo {title} {Evolution of magnetic structure driven by
  synthetic spin-orbit coupling in a two-component bose-hubbard model},}\
  }\href {\doibase 10.1103/PhysRevB.90.085117} {\bibfield  {journal} {\bibinfo
  {journal} {Phys. Rev. B}\ }\textbf {\bibinfo {volume} {90}},\ \bibinfo
  {pages} {085117} (\bibinfo {year} {2014}{\natexlab{b}})}\BibitemShut
  {NoStop}%
\bibitem [{\citenamefont {Piraud}\ \emph {et~al.}(2014)\citenamefont {Piraud},
  \citenamefont {Cai}, \citenamefont {McCulloch},\ and\ \citenamefont
  {Schollw\"ock}}]{mott_piraud}%
  \BibitemOpen
  \bibfield  {author} {\bibinfo {author} {\bibfnamefont {Marie}\ \bibnamefont
  {Piraud}}, \bibinfo {author} {\bibfnamefont {Zi}~\bibnamefont {Cai}},
  \bibinfo {author} {\bibfnamefont {Ian~P.}\ \bibnamefont {McCulloch}}, \ and\
  \bibinfo {author} {\bibfnamefont {Ulrich}\ \bibnamefont {Schollw\"ock}},\
  }\bibfield  {title} {\enquote {\bibinfo {title} {Quantum magnetism of bosons
  with synthetic gauge fields in one-dimensional optical lattices: A
  density-matrix renormalization-group study},}\ }\href {\doibase
  10.1103/PhysRevA.89.063618} {\bibfield  {journal} {\bibinfo  {journal} {Phys.
  Rev. A}\ }\textbf {\bibinfo {volume} {89}},\ \bibinfo {pages} {063618}
  (\bibinfo {year} {2014})}\BibitemShut {NoStop}%
\bibitem [{\citenamefont {Peotta}\ \emph {et~al.}(2014)\citenamefont {Peotta},
  \citenamefont {Mazza}, \citenamefont {Vicari}, \citenamefont {Polini},
  \citenamefont {Fazio},\ and\ \citenamefont {Rossini}}]{mott_peotta}%
  \BibitemOpen
  \bibfield  {author} {\bibinfo {author} {\bibfnamefont {Sebastiano}\
  \bibnamefont {Peotta}}, \bibinfo {author} {\bibfnamefont {Leonardo}\
  \bibnamefont {Mazza}}, \bibinfo {author} {\bibfnamefont {Ettore}\
  \bibnamefont {Vicari}}, \bibinfo {author} {\bibfnamefont {Marco}\
  \bibnamefont {Polini}}, \bibinfo {author} {\bibfnamefont {Rosario}\
  \bibnamefont {Fazio}}, \ and\ \bibinfo {author} {\bibfnamefont {Davide}\
  \bibnamefont {Rossini}},\ }\bibfield  {title} {\enquote {\bibinfo {title}
  {The xyz chain with dzyaloshinsky-moriya interactions: from
  spin-orbit-coupled lattice bosons to interacting kitaev chains},}\ }\href
  {http://stacks.iop.org/1742-5468/2014/i=9/a=P09005} {\bibfield  {journal}
  {\bibinfo  {journal} {Journal of Statistical Mechanics: Theory and
  Experiment}\ }\textbf {\bibinfo {volume} {2014}},\ \bibinfo {pages} {P09005}
  (\bibinfo {year} {2014})}\BibitemShut {NoStop}%
\bibitem [{\citenamefont {Pixley}\ \emph {et~al.}(2017)\citenamefont {Pixley},
  \citenamefont {Cole}, \citenamefont {Spielman}, \citenamefont {Rizzi},\ and\
  \citenamefont {Das~Sarma}}]{mott_pixley}%
  \BibitemOpen
  \bibfield  {author} {\bibinfo {author} {\bibfnamefont {J.~H.}\ \bibnamefont
  {Pixley}}, \bibinfo {author} {\bibfnamefont {William~S.}\ \bibnamefont
  {Cole}}, \bibinfo {author} {\bibfnamefont {I.~B.}\ \bibnamefont {Spielman}},
  \bibinfo {author} {\bibfnamefont {Matteo}\ \bibnamefont {Rizzi}}, \ and\
  \bibinfo {author} {\bibfnamefont {S.}~\bibnamefont {Das~Sarma}},\ }\bibfield
  {title} {\enquote {\bibinfo {title} {Strong-coupling phases of the
  spin-orbit-coupled spin-1 bose-hubbard chain: Odd-integer mott lobes and
  helical magnetic phases},}\ }\href {\doibase 10.1103/PhysRevA.96.043622}
  {\bibfield  {journal} {\bibinfo  {journal} {Phys. Rev. A}\ }\textbf {\bibinfo
  {volume} {96}},\ \bibinfo {pages} {043622} (\bibinfo {year}
  {2017})}\BibitemShut {NoStop}%
\bibitem [{\citenamefont {Fuchs}\ \emph {et~al.}(2005)\citenamefont {Fuchs},
  \citenamefont {Gangardt}, \citenamefont {Keilmann},\ and\ \citenamefont
  {Shlyapnikov}}]{Fuchs_PRL_2005}%
  \BibitemOpen
  \bibfield  {author} {\bibinfo {author} {\bibfnamefont {J.~N.}\ \bibnamefont
  {Fuchs}}, \bibinfo {author} {\bibfnamefont {D.~M.}\ \bibnamefont {Gangardt}},
  \bibinfo {author} {\bibfnamefont {T.}~\bibnamefont {Keilmann}}, \ and\
  \bibinfo {author} {\bibfnamefont {G.~V.}\ \bibnamefont {Shlyapnikov}},\
  }\bibfield  {title} {\enquote {\bibinfo {title} {Spin waves in a
  one-dimensional spinor bose gas},}\ }\href {\doibase
  10.1103/PhysRevLett.95.150402} {\bibfield  {journal} {\bibinfo  {journal}
  {Phys. Rev. Lett.}\ }\textbf {\bibinfo {volume} {95}},\ \bibinfo {pages}
  {150402} (\bibinfo {year} {2005})}\BibitemShut {NoStop}%
\bibitem [{ite()}]{itensor}%
  \BibitemOpen
  \href@noop {} {}\bibinfo {note} {Calculations performed using the ITensor C++
  library, \url{http://itensor.org/}}\BibitemShut {NoStop}%
\bibitem [{dmr()}]{dmrg_sym}%
  \BibitemOpen
  \href@noop {} {}\bibinfo {note} {In any finite system, the exact ground state
  is a ``cat state'' superposition of symmetry broken states with $\lr{s_3} =
  0$ identically. However, the DMRG truncation procedure favors low
  entanglement, and since the energy difference compared to the exact ground
  state is exponentially small in system size, the DMRG converges on one of the
  symmetry broken quasi-ground states with nonzero $\lr{s_3}$.}\BibitemShut
  {Stop}%
\bibitem [{\citenamefont {Yamamoto}(2015)}]{yamamoto}%
  \BibitemOpen
  \bibfield  {author} {\bibinfo {author} {\bibfnamefont {Naoki}\ \bibnamefont
  {Yamamoto}},\ }\bibfield  {title} {\enquote {\bibinfo {title} {Generalized
  bloch theorem and chiral transport phenomena},}\ }\href {\doibase
  10.1103/PhysRevD.92.085011} {\bibfield  {journal} {\bibinfo  {journal} {Phys.
  Rev. D}\ }\textbf {\bibinfo {volume} {92}},\ \bibinfo {pages} {085011}
  (\bibinfo {year} {2015})}\BibitemShut {NoStop}%
\bibitem [{\citenamefont {Alberton}\ \emph {et~al.}(2017)\citenamefont
  {Alberton}, \citenamefont {Ruhman}, \citenamefont {Berg},\ and\ \citenamefont
  {Altman}}]{altman_ruhman}%
  \BibitemOpen
  \bibfield  {author} {\bibinfo {author} {\bibfnamefont {Ori}\ \bibnamefont
  {Alberton}}, \bibinfo {author} {\bibfnamefont {Jonathan}\ \bibnamefont
  {Ruhman}}, \bibinfo {author} {\bibfnamefont {Erez}\ \bibnamefont {Berg}}, \
  and\ \bibinfo {author} {\bibfnamefont {Ehud}\ \bibnamefont {Altman}},\
  }\bibfield  {title} {\enquote {\bibinfo {title} {Fate of the one-dimensional
  ising quantum critical point coupled to a gapless boson},}\ }\href {\doibase
  10.1103/PhysRevB.95.075132} {\bibfield  {journal} {\bibinfo  {journal} {Phys.
  Rev. B}\ }\textbf {\bibinfo {volume} {95}},\ \bibinfo {pages} {075132}
  (\bibinfo {year} {2017})}\BibitemShut {NoStop}%
\bibitem [{\citenamefont {Dhar}\ \emph {et~al.}(2012)\citenamefont {Dhar},
  \citenamefont {Maji}, \citenamefont {Mishra}, \citenamefont {Pai},
  \citenamefont {Mukerjee},\ and\ \citenamefont {Paramekanti}}]{arun}%
  \BibitemOpen
  \bibfield  {author} {\bibinfo {author} {\bibfnamefont {Arya}\ \bibnamefont
  {Dhar}}, \bibinfo {author} {\bibfnamefont {Maheswar}\ \bibnamefont {Maji}},
  \bibinfo {author} {\bibfnamefont {Tapan}\ \bibnamefont {Mishra}}, \bibinfo
  {author} {\bibfnamefont {R.~V.}\ \bibnamefont {Pai}}, \bibinfo {author}
  {\bibfnamefont {Subroto}\ \bibnamefont {Mukerjee}}, \ and\ \bibinfo {author}
  {\bibfnamefont {Arun}\ \bibnamefont {Paramekanti}},\ }\bibfield  {title}
  {\enquote {\bibinfo {title} {Bose-hubbard model in a strong effective
  magnetic field: Emergence of a chiral mott insulator ground state},}\ }\href
  {\doibase 10.1103/PhysRevA.85.041602} {\bibfield  {journal} {\bibinfo
  {journal} {Phys. Rev. A}\ }\textbf {\bibinfo {volume} {85}},\ \bibinfo
  {pages} {041602} (\bibinfo {year} {2012})}\BibitemShut {NoStop}%
\bibitem [{\citenamefont {Celi}\ \emph {et~al.}(2014)\citenamefont {Celi},
  \citenamefont {Massignan}, \citenamefont {Ruseckas}, \citenamefont {Goldman},
  \citenamefont {Spielman}, \citenamefont {Juzeli\ifmmode~\bar{u}\else
  \={u}\fi{}nas},\ and\ \citenamefont {Lewenstein}}]{celi}%
  \BibitemOpen
  \bibfield  {author} {\bibinfo {author} {\bibfnamefont {A.}~\bibnamefont
  {Celi}}, \bibinfo {author} {\bibfnamefont {P.}~\bibnamefont {Massignan}},
  \bibinfo {author} {\bibfnamefont {J.}~\bibnamefont {Ruseckas}}, \bibinfo
  {author} {\bibfnamefont {N.}~\bibnamefont {Goldman}}, \bibinfo {author}
  {\bibfnamefont {I.~B.}\ \bibnamefont {Spielman}}, \bibinfo {author}
  {\bibfnamefont {G.}~\bibnamefont {Juzeli\ifmmode~\bar{u}\else
  \={u}\fi{}nas}}, \ and\ \bibinfo {author} {\bibfnamefont {M.}~\bibnamefont
  {Lewenstein}},\ }\bibfield  {title} {\enquote {\bibinfo {title} {Synthetic
  gauge fields in synthetic dimensions},}\ }\href {\doibase
  10.1103/PhysRevLett.112.043001} {\bibfield  {journal} {\bibinfo  {journal}
  {Phys. Rev. Lett.}\ }\textbf {\bibinfo {volume} {112}},\ \bibinfo {pages}
  {043001} (\bibinfo {year} {2014})}\BibitemShut {NoStop}%
\bibitem [{\citenamefont {Zheng}\ \emph {et~al.}(2014)\citenamefont {Zheng},
  \citenamefont {Liu}, \citenamefont {Miao}, \citenamefont {Chin},\ and\
  \citenamefont {Zhai}}]{zhai_quad}%
  \BibitemOpen
  \bibfield  {author} {\bibinfo {author} {\bibfnamefont {Wei}\ \bibnamefont
  {Zheng}}, \bibinfo {author} {\bibfnamefont {Boyang}\ \bibnamefont {Liu}},
  \bibinfo {author} {\bibfnamefont {Jiao}\ \bibnamefont {Miao}}, \bibinfo
  {author} {\bibfnamefont {Cheng}\ \bibnamefont {Chin}}, \ and\ \bibinfo
  {author} {\bibfnamefont {Hui}\ \bibnamefont {Zhai}},\ }\bibfield  {title}
  {\enquote {\bibinfo {title} {Strong interaction effects and criticality of
  bosons in shaken optical lattices},}\ }\href {\doibase
  10.1103/PhysRevLett.113.155303} {\bibfield  {journal} {\bibinfo  {journal}
  {Phys. Rev. Lett.}\ }\textbf {\bibinfo {volume} {113}},\ \bibinfo {pages}
  {155303} (\bibinfo {year} {2014})}\BibitemShut {NoStop}%
\bibitem [{\citenamefont {{Po}}\ and\ \citenamefont
  {{Zhou}}(2015)}]{zhou_quad2}%
  \BibitemOpen
  \bibfield  {author} {\bibinfo {author} {\bibfnamefont {H.~C.}\ \bibnamefont
  {{Po}}}\ and\ \bibinfo {author} {\bibfnamefont {Q.}~\bibnamefont {{Zhou}}},\
  }\bibfield  {title} {\enquote {\bibinfo {title} {{A two-dimensional algebraic
  quantum liquid produced by an atomic simulator of the quantum Lifshitz
  model}},}\ }\href {\doibase 10.1038/ncomms9012} {\bibfield  {journal}
  {\bibinfo  {journal} {Nat. Commun.}\ }\textbf {\bibinfo {volume} {6}},\
  \bibinfo {eid} {8012} (\bibinfo {year} {2015})}\BibitemShut {NoStop}%
\bibitem [{\citenamefont {Radi\ifmmode~\acute{c}\else \'{c}\fi{}}\ \emph
  {et~al.}(2015)\citenamefont {Radi\ifmmode~\acute{c}\else \'{c}\fi{}},
  \citenamefont {Natu},\ and\ \citenamefont {Galitski}}]{radic_quad}%
  \BibitemOpen
  \bibfield  {author} {\bibinfo {author} {\bibfnamefont {Juraj}\ \bibnamefont
  {Radi\ifmmode~\acute{c}\else \'{c}\fi{}}}, \bibinfo {author} {\bibfnamefont
  {Stefan~S.}\ \bibnamefont {Natu}}, \ and\ \bibinfo {author} {\bibfnamefont
  {Victor}\ \bibnamefont {Galitski}},\ }\bibfield  {title} {\enquote {\bibinfo
  {title} {Strong correlation effects in a two-dimensional bose gas with
  quartic dispersion},}\ }\href {\doibase 10.1103/PhysRevA.91.063634}
  {\bibfield  {journal} {\bibinfo  {journal} {Phys. Rev. A}\ }\textbf {\bibinfo
  {volume} {91}},\ \bibinfo {pages} {063634} (\bibinfo {year}
  {2015})}\BibitemShut {NoStop}%
\bibitem [{\citenamefont {Wu}\ \emph {et~al.}(2017)\citenamefont {Wu},
  \citenamefont {Zhou},\ and\ \citenamefont {Wu}}]{wu_quad}%
  \BibitemOpen
  \bibfield  {author} {\bibinfo {author} {\bibfnamefont {Jianda}\ \bibnamefont
  {Wu}}, \bibinfo {author} {\bibfnamefont {Fei}\ \bibnamefont {Zhou}}, \ and\
  \bibinfo {author} {\bibfnamefont {Congjun}\ \bibnamefont {Wu}},\ }\bibfield
  {title} {\enquote {\bibinfo {title} {Quantum criticality of bosonic systems
  with the lifshitz dispersion},}\ }\href {\doibase 10.1103/PhysRevB.96.085140}
  {\bibfield  {journal} {\bibinfo  {journal} {Phys. Rev. B}\ }\textbf {\bibinfo
  {volume} {96}},\ \bibinfo {pages} {085140} (\bibinfo {year}
  {2017})}\BibitemShut {NoStop}%
\bibitem [{\citenamefont {Balents}\ and\ \citenamefont
  {Starykh}(2016)}]{balents_quad}%
  \BibitemOpen
  \bibfield  {author} {\bibinfo {author} {\bibfnamefont {Leon}\ \bibnamefont
  {Balents}}\ and\ \bibinfo {author} {\bibfnamefont {Oleg~A.}\ \bibnamefont
  {Starykh}},\ }\bibfield  {title} {\enquote {\bibinfo {title} {Quantum
  lifshitz field theory of a frustrated ferromagnet},}\ }\href {\doibase
  10.1103/PhysRevLett.116.177201} {\bibfield  {journal} {\bibinfo  {journal}
  {Phys. Rev. Lett.}\ }\textbf {\bibinfo {volume} {116}},\ \bibinfo {pages}
  {177201} (\bibinfo {year} {2016})}\BibitemShut {NoStop}%
\bibitem [{\citenamefont {Hals}(2017)}]{hals_schm}%
  \BibitemOpen
  \bibfield  {author} {\bibinfo {author} {\bibfnamefont {Kjetil M.~D.}\
  \bibnamefont {Hals}},\ }\bibfield  {title} {\enquote {\bibinfo {title}
  {Magnetoelectric coupling in superconductor-helimagnet heterostructures},}\
  }\href {\doibase 10.1103/PhysRevB.95.134504} {\bibfield  {journal} {\bibinfo
  {journal} {Phys. Rev. B}\ }\textbf {\bibinfo {volume} {95}},\ \bibinfo
  {pages} {134504} (\bibinfo {year} {2017})}\BibitemShut {NoStop}%
\bibitem [{\citenamefont {{Greiner}}\ \emph {et~al.}(2002)\citenamefont
  {{Greiner}}, \citenamefont {{Mandel}}, \citenamefont {{Esslinger}},
  \citenamefont {{H{\"a}nsch}},\ and\ \citenamefont {{Bloch}}}]{greiner}%
  \BibitemOpen
  \bibfield  {author} {\bibinfo {author} {\bibfnamefont {M.}~\bibnamefont
  {{Greiner}}}, \bibinfo {author} {\bibfnamefont {O.}~\bibnamefont {{Mandel}}},
  \bibinfo {author} {\bibfnamefont {T.}~\bibnamefont {{Esslinger}}}, \bibinfo
  {author} {\bibfnamefont {T.~W.}\ \bibnamefont {{H{\"a}nsch}}}, \ and\
  \bibinfo {author} {\bibfnamefont {I.}~\bibnamefont {{Bloch}}},\ }\bibfield
  {title} {\enquote {\bibinfo {title} {{Quantum phase transition from a
  superfluid to a Mott insulator in a gas of ultracold atoms}},}\ }\href
  {\doibase 10.1038/415039a} {\bibfield  {journal} {\bibinfo  {journal} {\nat}\
  }\textbf {\bibinfo {volume} {415}},\ \bibinfo {pages} {39--44} (\bibinfo
  {year} {2002})}\BibitemShut {NoStop}%
\end{thebibliography}


%%%%%%%%%%%%%%%%%%%%%%%%%%%%%%%%%%%%%%%%%%%%%%%%%%

%\subsection*{Author contributions}
%I.B.S. and J.D.S. concieved the project; W.S.C., Y.A., and J.D.S. carried out the field theory calculations; J.L. and K.W.M. carried out the numerical simulations; W.S.C., J.L., and J.D.S. wrote the initial manuscript. All authors contributed to the analysis of results and reviewing the manuscript.

\end{document}